\input amstex 
\documentstyle{amsppt}
\input bull-ppt

\let\sl\it
\let\frak\germ

\define\gr{renormalization-group }
\define\nspace{\lineskip=1pt\baselineskip=12pt%
\lineskiplimit=0pt}

\define\references#1{\bigskip\par\centerline{\bf 
References}\medskip
     \parindent=#1pt\nspace}
\define\Ref#1{\par\smallskip\hang\indent\llap{\hbox 
to\parindent
     {#1\hfil\enspace}}\ignorespaces}


\topmatter
\cvol{30}
\cvolyear{1994}
\cmonth{January}
\cyear{1994}
\cvolno{1}
\cpgs{1-61}
\title Conformal invariance in two-dimensional percolation 
\endtitle
\author Robert Langlands, Philippe Pouliot, and  Yvan 
Saint-Aubin \endauthor
\address School of Mathematics,
Institute for Advanced Study,
Princeton, New Jersey 08540\endaddress
\ml rpl\@math.ias.edu\endml
\address D\'epartement de Physique and  
Centre de Recherches Math\'ematiques,
Universit\'e de Montr\'eal,
C.P.~6128A, Montr\'eal, Qu\'ebec, 
Canada H3C 3J7\endaddress
\cu Department of Physics,
Rutgers University,
New Brunswick, New Jersey 08855 \endcu
\ml pouliot\@physics.rutgers.edu\endml
\address Centre de Recherches Math\'ematiques and 
D\'epartement de Math\'ematiques
et de Statistique, 
Universit\'e de Montr\'eal, 
C.P.~6128A, Montr\'eal, Qu\'ebec,
Canada H3C 3J7\endaddress
\ml saint\@crm.umontreal.ca\endml
\date October 4, 1992 and, in revised form,
December 8, 1992, and July 20, 1993\enddate
\thanks A first version of part of the material of this
paper was presented by the first author as part of the
AMS Colloquium lectures in Baltimore in January 
1992\endthanks
\thanks The third author was supported 
in part by NSERC Canada and the Fonds FCAR pour l'aide
et le soutien \`a la recherche (Qu\'ebec)\endthanks
\endtopmatter

\document  
\heading Contents\endheading

1. Introduction.

2. The hypotheses of universality and conformal invariance.

\ \ 2.1 Basic results and questions in percolation.

\ \ 2.2 Universality and the renormalization group.

\ \ 2.3 Crossing probabilities.

\ \ 2.4 The two hypotheses.

\ \ 2.5 More critical indices for percolation.

\ \ 2.6 Conformally invariant fields and percolation.

3. The experiments.

\ \ 3.1 Experimental procedure.

\ \ 3.2 Experimental verification of Cardy's formula.

\ \ 3.3 Parallelograms.

\ \ 3.4 Striated models.

\ \ 3.5 Exterior domains.

\ \ 3.6 Branched percolation.

\ \ 3.7 Percolation on compact Riemann surfaces.

\heading 1. Introduction\endheading

 The word {\it percolation},
borrowed from
the Latin, refers to the seeping or
oozing of a liquid through a porous medium,
usually to be strained. In this and related senses
it has been in use since the seventeenth century.
It was introduced more recently into mathematics
by S.~R.~Broadbent and J.~M.~Hammersley (\cite {BH})
and is a branch of probability theory that
is especially close to statistical
mechanics. 
Broadbent and Hammersley distinguish between
two types of spreading of a fluid through a
medium, or between two aspects of
the probabilistic models of such processes:
{\it diffusion} processes, in which the
random mechanism is ascribed to the fluid;
and {\it percolation} processes, in which
it is ascribed to the medium.

 A percolation process typically depends on one
or more probabilistic parameters. For example, if
molecules of a gas are absorbed at the
surface of a porous solid (as in a gas mask)
then their ability to penetrate the solid
depends on the sizes of the pores in it
and their positions, both conceived
to be distributed in some random manner. A 
simple mathematical
model of such a process is often defined
by taking the pores to be distributed
in some regular manner
(that could be determined by
a periodic graph), and to be open (thus very large) or
closed (thus smaller than the molecules) with 
probabilities $p$
and $1-p$.
As $p$ increases the probability of 
deeper penetration of the gas into the interior of
the solid  grows.

 There is often a {\it critical} threshold
for the probability at which the behavior
changes abruptly --- below which the penetration
is only superficial, and above which
it is infinitely deep. Such critical behavior
is a very simple analogue of similar behavior
in thermodynamics and statistical mechanics
that is of great theoretical and experimental,
as well as mathematical, interest. 
Since the critical behavior
manifested in percolation shares many characteristics
with that of more complex systems and models,
percolation has attracted wide interest (\cite{G,K})
among physicists and mathematicians as one of the
simplest cases in which various striking
features of critical behavior, especially
{\it scaling} and
{\it universality}, appear. These two terms are 
central to this paper, and will be discussed
more at length below. Scaling refers,
in essence, to
the frequent appearance of simple power laws.
The exponent in these laws is often the same
for quite different
materials and models, and this is
called
universality.

 The immediate purpose of the paper was neither to review 
the
basic definitions of percolation theory nor to rehearse the
general 
physical notions of universality
and renormalization
(an important technique to be described in Part Two).
It was rather to describe
as concretely as possible,
although in hypothetical form, the geometric aspects of 
universality, especially conformal
invariance, in the context of percolation,
and to present the numerical results that support the
hypotheses. On the other hand,
one ulterior purpose is to
draw the attention of mathematicians to
the mathematical
problems posed by the physical
notions. Some precise basic
definitions are necessary simply to orient the
reader. Moreover a brief description of scaling and 
universality on
the one hand and of renormalization on the other
is also essential
in order to establish  
their physical importance and
to clarify their mathematical content. 

 These matters are all treated in Part Two.
Since one of its purposes is to 
orient ourselves and other
inexperienced mathematicians
with respect to the physical
background, we have not shrunk
from the occasional doubtful utterance
that shed, for us at least,
some light in an obscure corner. We urge the reader
to be especially circumspect while reading
\S 2.2. That we are dealing there with material
with which none of us has had first-hand experience
is not the least of the reasons, but it is also not
the only one.

 The first paragraph of Part Two
is deliberately stark. We hope that the
content of the questions posed there is clear; their
depth cannot be at this stage. 
They are central and inaccessible,
but as problems
they are the source of the
hypotheses of  \S 2.4 and the
experiments described in Part Three.

 The attention given
to \S 2.2 will depend on the reader's familiarity
with the physical concepts used.
Many are fairly close to
everyday experience, but 
there are also deep ideas with a long history
compressed into single phrases. Fortunately
the section can be skipped completely, and
those with no experience
with the concepts
can pass directly,
or at least quickly, to \S 2.3 and \S 2.4
which are prerequisites
to Part Three. \S 2.2 is not. Nor
are the final two paragraphs of Part Two.
\S 2.5 is an appendix, in the context of
percolation, to \S 2.2. The material in \S 2.6 is
especially difficult, but especially important because
it illustrates the power of the methods of
conformal field theory for making {\it analytic}
predictions. These appear to be far less accessible
to rigorous mathematical demonstration,
and perhaps deeper, than more familiar
{\it geometric} predictions. The ideas of 
\S 2.6 are due to
Cardy, and appear in a sequence of papers. In spite
of their lack of rigor, they appear to be of great 
potential,
and our purpose is simply to present them in the most
accessible form we could manage.

 Since only the statements
of the hypotheses are strict prerequisites for it,
Part Three, far more elementary than
Part Two, can be
read without a thorough understanding of the preceding part.
By the same token, Part Three can be taken as 
nothing more than an illustration
of what happens when mathematicians take the
physical ideas of Part Two seriously, and Part
Two can
be read without reference to it. 

 After the discussion of the general experimental 
procedure in
\S 3.1, the description of the experiments begins. 
It is, of course, the experiments that give substance to
the paper, in which nothing is proved
mathematically. \S 3.2 offers
a table of approximate, but
statistically very precise results obtained by simulation
that serve two purposes: a verification with better
data than those of \cite{U} of the formula
of Cardy in \S2.6; construction of a collection
of data with which the less precise data
of the following sections may be compared.

 The numerical investigation of conformal invariance
is begun in \S3.3. The data of \S 3.2 are for
rectangles. The interior of every parallelogram is 
conformally
equivalent to the interior of an
appropriate rectangle,
and the conformal mapping is 
uniquely
determined if it is insisted that vertices
be taken to vertices.
Moreover the aspect ratio $r$ of the rectangle
(the quotient of the lengths of neighboring sides)
is all but uniquely determined. The only possibility
is that $r$ be replaced by $1/r$. Thus a natural
first comparison to establish conformal 
invariance is to compare data for parallelograms
with the standard data of \S 3.2 for rectangles.
This is done in \S 3.3.

The notion of universality of \S 2.4 is not that
of \S 2.2, but closely related to it; and as
remarked
in \cite{U} it is difficult to determine to what
extent it was accepted in the community of specialists.
It has certainly not been exploited. Specialists
are not inclined to doubt it
when questioned closely
and it has been tested in a restricted
form in \cite{U}. In \S3.4, we content ourselves with
a single example of the general hypothesis, whose
purpose is principally to exhibit an example
in which all symmetries are violated, and to show
how to make calculations for it.

 The final three sections in Part Three are a 
more adventurous pursuit of the consequences
of conformal invariance of percolation.
We define percolation on a variety of 
Riemann surfaces: unbounded planar domains; branched
coverings of bounded planar domains;
and then on branched coverings
of the Riemann sphere. We stop there, but we
could have gone farther. The principle has
certainly become clear. In each case,
we take an example and verify conformal
invariance for it, but for reasons that
we explain the precision with which we verify this 
invariance
decreases. Thus the numerical evidence for
conformal invariance in the generality it is
finally conceived is not so good as
it could be with more painstaking
experiments, but even those performed
took considerable time, and provide
evidence that is positive,
and in our view convincing.
Our aim was less to achieve
great precision than to assure ourselves 
that even bold forms of the
hypothesis of conformal invariance
stood a good chance of being valid. Although
further precision is certainly desirable,
it seems to us that the search for proofs
can begin with some moral certainty that the
general assertions implicit
(the reader will have no difficulty
in making them explicit) in the last three
sections are valid.

 As far as we have been able to determine (with the help 
provided by
A. Mortensen of the Department of Material Sciences at MIT)
the study of critical behavior and universality
in percolation is of much less
practical than theoretical importance. 
The paper \cite{M} of MacLachlan {\it et al} 
and that of Wong \cite{W} suggest that in such practical
applications of percolation processes as 
the study of composite materials or the
porosity of rocks the interest is less in quantities
similar to that of the theorem of \S 2.1
that change abruptly at the critical threshold
than in quantities such as conductivity or
permeability that change continuously,
although with an infinite derivative,
across this threshold. The critical indices
of this paper are important in so far as they influence
the equations governing these quantities, but the
principal {\it practical} problem is perhaps to
reduce, geometrically or otherwise, the critical threshold,
for this means incorporating less of a perhaps
expensive additive in an inexpensive matrix.
Our concerns are theoretical and mathematical.

\heading 2. The hypotheses of universality and conformal 
invariance\endheading

\subheading{\num{2.1.} Basic results and questions in 
percolation}
 A standard model of percolation is that attached to sites 
on
a square lattice.  Let $L$ be the graph (embedded in $\Bbb 
R^d$)
whose set of vertices or {\it sites\/} is the set of
integral points $\Bbb Z^d$ and whose edges or {\it 
bonds\/} join
all pairs of nearest neighbors.
Each site can be in one of two {\it states}. It
can be {\it open}
and then we assign it the value $1$,
or it can be {\it closed},
and be assigned the value $0$.
A {\it configuration} is obtained by specifying
which sites are open and which are closed.
Clearly the set $X$ of all configurations is
$$
\prod_{\Bbb Z^d}\{0,1\},
$$
the set of functions from
$\Bbb Z^d$ to $\{0,1\}$.
A site $s$ is open for a configuration
if the corresponding function
takes the value $1$ at $s$. If $0\leq p\leq 1$
then we associate to $p$ the probability
on $\{0,1\}$ that assigns the probability $p$ to
$1$, and
introduce the product of these probabilities 
on the set of all configurations. Each site 
can then be regarded as
an independent random variable assuming two possible
values $0$ or $1$. We refer to
the set $X$ with this probability measure as
the model $M_0$ of percolation.

 For many purposes it is convenient
to work
not with the full graph $L$
but with the sites 
$$
      \{(i,j)\vert 1\leq i,j \leq n\}
$$
in a square $S_n$ of side $n$ and the bonds connecting
them. If 
$$
  X_n=\prod_{S_n}\{0,1\},
$$
then configurations $x\in X_n$ are determined by
fixing a state for each site
in $S_n$. The
probability $\pi(x)$
of $x$ is equal to $p^k(1-p)^l$ if $k$ sites are open for 
$x$,
and $l=n^2-k$ are closed.
A typical configuration $x$ is shown in Figure 2.1a, in 
which
open sites appear as black dots and closed sites are
white.


\fighere{19 pc}\caption{Figure 2.1\rm a.  Configurations 
on the
square cube $S_{16}$ for percolation by sites.}
\fighere{19 pc}\caption{Figure 2.1\rm b. Configurations on 
the
square cube $S_{16}$ for percolation by bonds. Both (a) 
and (b)
have a horizontal crossing but no vertical one.}

 There are many different events
in $X$ or $X_n$ whose
probabilities are of interest in the study of
percolation. We shall return to them
in \S 2.3. For now,
in order to put
the questions in stark simplicity, we concentrate
on a very special probability  $\pi_h$, that of a 
horizontal crossing.
Consider the configuration $x$ on $S_{16}$ of Figure 2.1a. 
This configuration admits a horizontal crossing in the sense
that it is possible to pass from the
left side of the square to the
right one by moving
repeatedly from one site open for $x$ to another open site
joined to it by a bond, thus to an open immediate neighbor.
It does not, however, admit a vertical crossing.
The probability $\pi_h^n(p)$ of a horizontal crossing
is the sum  of the probabilities $\pi(x)$,
taken over all configurations $x\in X_n$ on $S_n$ 
that admit a horizontal crossing.


\fg{16 pc}\caption{Figure 2.1\rm c. The curves 
$\pi_h^n(p)$ for
$n=2,4,8,16, 32, 64$, and 128. Larger slopes around $p_c$
correspond to larger values of $n$.}
\endfg

 The probability $\pi_h^n(p)$ clearly increases from $0$
to $1$ as $p$ does. Its behavior 
with respect to $n$ is revealed by Figure 2.1c,
in which the graph  of the function
$\pi_h^n$ is given for
$n=2,4,8,16,32,64,128$. It appears to be approaching
a step function; this
is confirmed by
the first two statements of the following theorem, whose
original proof takes up most of the book  \cite {K}
of Kesten.  A full account of
the contributions of earlier authors can be found
there. A more recent
proof can be found in \cite {AB}.
\proclaim{Theorem } There exists a unique critical
probability $0<p_c<1$ such that:
\roster
\item
for $p<p_c$,
$$
    \lim_{n\rightarrow\infty}\pi_h^n(p)=0;
$$
\item
for $p>p_c$,
$$
    \lim_{n\rightarrow\infty}\pi_h^n(p)=1;
$$
\item
for $p=p_c$,
$$
   0<\liminf_{n\rightarrow\infty}\pi_h^n(p)\leq
     \limsup_{n\rightarrow\infty}\pi_h^n(p)<1.
$$
\endroster
\endproclaim

 In spite of the difficulty and importance of the
theorem, it has an obvious defect for it does
not answer the question that immediately springs
to mind upon reading the final statement.
\dfn{Question 1} Does 
$$
    \lim_{n\rightarrow\infty}\pi_h^n(p)
$$
exist for $p=p_c$?
\enddfn

 The numerical evidence leaves no doubt that the limit,
which we denote $\pi_h$, exists. A second question, far more
subtle, is also strongly suggested by the numerical
data. Consider the derivative $A_n$ of $\pi_h^n(p)$ with 
respect
to $p$ at $p=p_c$. If Figure 2.1c does not deceive
then $A_n$ increases with $n$ and approaches infinity.
\dfn{Question 2}
Does there exist a  positive real number $\nu$ such that
$$    
\lim_{n\rightarrow\infty}\frac{A_n}{n^{\tfrac1\nu}}\tag 
2.1a
$$
exists and is different from $0$?
\enddfn
 
 This is a simple
example of a scaling law, a notion 
that will be explained more generally in the next section.

 The two questions, as well as
the theorem, have been formulated for
the specific model $M_0$, but there are many other 
possible models. For example, in dimension two
the lattice $\Bbb Z^2$ 
can be replaced by a triangular (or hexagonal)
lattice in which each site has 6 (3) nearest neighbors. 
Percolation by sites can also be
replaced with percolation by bonds. 
In bond percolation all sites are open and 
it is the bonds that are open with
probability $p$. A 
configuration on $S_{16}$ is shown in Figure 2.1b.
The definitions introduced for site percolation
on $M_0$
are applicable to these new models. The configuration in
the figure admits a horizontal crossing but no vertical
crossing.
One can also study percolation on more general planar 
graphs,
allowing in addition both sites and bonds to be open or 
closed,
and probabilities that depend on the type of bond or site. 
We could, for example, in bond percolation on a square 
lattice
permit the horizontal and vertical bonds to be open with
different probabilities $p_h$ and $p_v$.
The variations are endless, but
for all models within a large class, the
theorem,
in an appropriate form, remains valid,
and the questions appear to continue to have
an affirmative response. The critical probabilities
vary from model to model, but the evidence strongly
suggests that yet another, a third, question has an 
affirmative
answer.
\dfn{Question 3}
Is the value $\nu$ 
independent of the model?
\enddfn

 The number $\nu$ is known as a critical index
and its independence of the model is known as universality.
For reasons not germane to this paper $\nu$ is generally
believed to be equal to the rational number $\tfrac43$
for the models of percolation in two dimensions that we
study here.

 The first, obvious advantage of percolation models
is the facility with which $\nu$ can be introduced.
In statistical mechanics singular behavior
of quantities such as
specific heat or magnetic susceptibility
is also described by  
critical indices, to be discussed
in the next paragraph, whose constancy within large
classes of models,
thus their universality, is
well established within the limits
of experimental observation.
Although its sources are not 
understood, there is a very powerful method,
the {\it renormalization group},
for analyzing critical behavior, but the problem
of understanding the mechanism that allows
the geometry to predominate and to efface the 
details of the interactions 
and, as a consequence, to render renormalization
so effective remains. 
The missing insight can be regarded as
physical or mathematical; it is not
a question of adding rigor to arguments that
are otherwise persuasive. 
There are none.

 The renormalization group was taken, as its
name suggests, into statistical mechanics
from the theory of quantum fields, and has therefore
a conceptually very difficult history with
which we are not concerned, although some attempt
will be made during the course of the paper
to give the phrase some meaning to the reader.
It should then be clear to him that, contrary
to the first impression, the three questions
are not at an ever increasing level of difficulty,
so that an earlier one must be answered before
a later one can be posed. They must rather be 
answered simultaneously.

 With this in mind, our purpose, in \cite{L1,L2} and 
\cite{U}, has been
to introduce objects that 
deserve to be called renormalizations, but that are at
the same time concrete, elementary mathematical
objects amenable to rigorous mathematical investigation.

 What is introduced in \cite{L1,L2} is a sequence of
continuous
transformations of finite-dimensional
spaces. They are briefly reviewed in 
\S 2.3.
To relate these objects to renormalization  
requires hypotheses whose validity
was not universally accepted.
To assure ourselves that the definitions
were well-founded
we examined crossing probabilities like $\pi_h$
for various models of percolation in \cite{U}. 
Conversations with Michael
Aizenman 
after the data were in hand
greatly clarified for us their nature. In particular 
he suggested that these crossing probabilities would be 
conformally invariant.

 Subsequent conversations with other mathematicians 
persuaded us that 
with the appearance of conformal invariance percolation 
becomes a topic
that appeals to a broader audience than mathematical 
physicists
and probabilists. For example, a remark of Israel Gelfand, 
for 
which we are grateful, led to the
examination of conformal invariance on compact Riemann 
surfaces. Since
proofs of conformal invariance will likely have to wait upon
proofs of universality
for percolation,
and these, even  if the ideas of \cite{L1,L2}
have some validity, will in all likelihood be slow in 
coming, we decided
to present the numerical evidence for 
conformal invariance and its consequences
in a form that emphasizes its mathematical appeal, and 
this is the
primary purpose
of the
present 
paper. No theorems are proved or implied.

 As promised, 
we preface the numerical
results with an explanation, tailored to
our concerns, of the terms,
universality and renormalization,
just invoked. Before beginning, we would like
to express our thanks to Michael Aizenman and to Thomas 
Spencer
for their encouragement.

\subheading{\num{2.2.} Universality and the 
renormalization group}
 Statistical mechanics and the closely related subject
of thermodynamics deal, to some extent, with objects
familiar to all of us: gases, liquids, and solids;
or magnets in magnetic fields. It comes,
therefore, as somewhat of a shock to learn
that these substances are not so familiar
as we might think. Water vapor, water, and ice
and the transitions between them are matters of daily
experience, and phase diagrams like Figure 2.2a
frequently met. 


\fg{24 pc}\caption{Figure 2.2\rm a. Qualitative phase 
diagram
for water.}
\endfg
\fg{16 pc}\caption{Figure 2.2\rm b. Qualitative phase 
diagram
for a ferro\-magnet.}
\endfg

 They are not usually drawn to scale nor do we ask ourselves
which region or values of the pressure and temperature
are accessible under normal conditions. Temperatures
between $-20^\circ\ {\text C}$ and $100^\circ\ {\text C}$, 
the boiling point of water,
are the most common, except under incendiary conditions.
Because of the phenomenon
of partial pressure, more
familiar
to us as the numerator in the {\it humidity},
only the pressure of the water vapor in the ambient
air affects the rate of evaporation or thaw,
so that
the pertinent range of pressures is from $1. \text{\rm\ 
atm}$
all the way down to $0. \text {\rm\ atm}$. Thus, even though
the triple point $A$ in Figure 2.2a is at
$(P,T)=(0.006 \text{ \rm\ atm}, 0.^\circ\ {\text C}$), ice 
does melt
on the surface of ponds and puddles.  

 On the other hand, the point $B$, the critical point
in the technical sense, is at 
$(P_c,T_c)=
(218.\text{\rm \ atm},341.^\circ \text{C})$, so that
no diagram drawn to scale could include the two points.
The pressure is that found more than two kilometers 
under the ocean surface, not a familiar location,
and certainly not one in which we might try to boil water.

 Thus the 
phenomena associated with the critical point, and it is 
for them
that universality is pertinent, are not those
associated to the
transition from water to ice or from water to water vapor.
They are of a different nature. If at
a fixed temperature  $T$  below
$T_c$ we continuously increase the pressure
(or reduce the volume)
on a closed container of water vapor then,
when the pressure is such that $(P,T)$ lies on the curve
$C$, it will start to condense and we will be able
to continue to reduce the volume without changing
the pressure until there is no vapor left.
At this point, continued reduction of the
volume will increase the pressure, or more
kinesthetically, continued increase
of the pressure will reduce the volume,
which will have decreased considerably.
It is best to imagine
the transition occurring in the
absence of a gravitational field,
so that the difference of density does not cause,
in the familiar way,
the liquid to precipitate out. Rather a kind of slush is
formed during the transition, pockets of liquid
in the ambient vapor, or pockets
of air in the ambient liquid.

 At the point on the curve, where the volume,
and therefore the density $\rho$, changes
without any change in the pressure, the
isothermal compressibility 
$$
     K_T=\frac1\rho\left(\frac{\partial \rho}{\partial 
P}\right)_T\tag 2.2a
$$
is of course infinite.
Above $T_c$ the curve $C$ has terminated
and there is no transition from vapor to liquid,
rather there is simply a fluid that is gradually
becoming denser with the increase in pressure.
In particular, at no point does $K_T$   
become infinite.
 
 If the pressure is increased in the same way at 
$T=T_c$, the behavior can be expected
to mimic both that at $T<T_c$ and at
$T>T_c$. The curve given by setting $T$ equal
to a constant $T_c$ and letting $P$ vary
could be replaced by other curves passing through
the critical point,
but it is better to work
with a fixed, simply defined curve. 
We observe, anticipating a later section,
that 
the critical behavior of percolation, in which there
is only one free parameter, the probability,
is to be compared with the behavior 
along such a curve.

 The fluid, whether a liquid or a gas, is composed
of molecules that are subject to
thermal fluctuations, so that the density
is only defined for statistically significant
aggregates of molecules. Away from the curve
$C$ a few molecules suffice 
(cf \cite {P}) so that the {\it normal}
or {\it bulk} state is achieved in aggregates
occupying a region whose size
usually is of the order of
a few molecular diameters, thus
of the order of $3\times 10^{-10}\text{\rm m}$.
On the curve itself, a bulk state is a mixture,
with regions, gaseous or liquid,
visible to the naked eye, whose size, in terms
of molecular diameters, is therefore
effectively infinite. 

 The size required in order for quantities
like the density to be defined is
usually, for statistical reasons, referred to
as the correlation length and denoted by $\xi$.
It depends on the pressure and the temperature,
$\xi=\xi(P,T)$, and becomes infinite 
at the critical point $B$
because, for the reasons given, it is
infinite along the curve.
Thus the scale on which the thermal fluctuations
occur
grows as the critical point
is approached, eventually reaching 
and surpassing the wavelength
of visible light, about  $5\times 10^{-7}\text{\rm m}$.

 Although our initial discussion is for water, because
it is so common, it may not be, as the following citation 
suggests,
the best substance with which to conduct experiments
around the critical point. For reasons described clearly 
and simply
in \cite {S}, they are very difficult. 

 The optical phenomena, known as 
{\it critical opalescence}, that result from
the increase in correlation length are quite colorful
and very famous. Unfortunately, the best photographs
and slides
have never, to our knowledge, been published. We refer the 
reader to the cover
of the June 10, 1968 issue of 
Chemical and Engineering News for the only color
reproduction known to us. It would be useful, and
would clear up many
common misconceptions, if photographs illustrating
the brownish-orange stage of
Michael Fisher's description
of critical opalescence for carbon dioxide in \cite{F2}
were published:
{\smallskip\narrower\narrower\noindent
``if the carbon dioxide, which is quite transparent in the 
visible region
of the spectrum, is illuminated from the side, one 
observes a strong
intensity of scattered light. This has a bluish tinge when 
viewed
normal to the direction of illumination, but has a 
brownish-orange
streaky appearance, like a sunset on a smoggy day, when 
viewed from the
forward direction (i.e., with the opalescent fluid 
illuminated
from behind). Finally, when the temperature is raised a 
further few tenths
of a degree, the opalescence disappears and the fluid 
becomes completely 
clear again.''\medskip}

 We review as briefly as we can, in a form
suitable for mathematical consumption, the conceptual
conclusions from the experiments.
Our discussion, which begins with scaling and universality,
is taken from \cite{F1} and the companion survey \cite {H} 
of
experimental results. The notion of renormalization had
not appeared in the theory at this stage. We stress at
the outset that scaling is one conclusion from
the experimental evidence, and universality a second.
Renormalization is, for the moment, a largely heuristic
mathematical argument to explain them both.

 Although the details of the phase diagram varies
from substance to substance, it remains qualitatively
the same, and the 
behavior of the correlation length $\xi$
does not change.
As far as can be determined it
behaves near the critical point like a power of the
distance $\rho$ from the critical point
$$
   \xi\sim  \rho^{-\nu}. \tag 2.2b
$$
On the curve defined by setting $T$ equal to $T_c$, the 
parameter
$\rho$ is $\vert P-P_c\vert$; on that defined by setting
$P=P_c$ it is $\vert T-T_c\vert$.

 The equation (2.2b)
is another instance of scaling that can immediately be 
compared
with that of (2.1a). The correlation length is
the size of the sample that is necessary for local
statistical irregularities to be disregarded, so that
the substance is in a normal or bulk state. For percolation,
when the parameter $p$ is not equal to $p_c$ this is the 
size 
at which the conclusions of the first or the second part 
of the theorem take effect, thus for which $\pi_h^n(p)$
is very close either to $0$ or to $1$. Since $A_n$
is the derivative at $p_c$ this requires,
according to the third part of the theorem, that
the absolute
value of $A_n(p-p_c)$ be a number $B$ bounded away from $0$.
The smallest $n$ at which this occurs is a candidate
for the correlation length.
The two relations $\xi=n$ and
$$
     A_n\vert p-p_c\vert=B,
$$
together with (2.2b) yield
$$
   \vert p-p_c\vert^{-1} \sim A_n\sim 
n^{\tfrac1\nu}=\xi^{\tfrac1\nu},
$$
or $\xi\sim \vert p-p_c\vert^{-\nu}$.
   
 Although the critical exponent $\nu$ is the obvious
one for percolation,
and fundamental in general, it is one of the most difficult
to
measure experimentally.
For the liquid-gas transition in pure fluids
Fisher asserts
(\cite {F2}) that it has
a value in the range $0.55$ to $0.70$, implicitly
suggesting that its value is independent of the fluid,
thus universal. Since this is certainly not the value
$\tfrac43$ that appears to
be correct for percolation, there 
must certainly be more than one universality class.

 Although the phenomena of universality and scaling
were discovered prior to the introduction of the
renormalization group, it is easier to persuade
the mathematical reader of the delicacy
of the notion of universality classes
into which real substances and models are supposed
to fall, if it is explained immediately that
they are expected to correspond to
the stable manifolds of unstable
fixed points of the renormalization group
transformation that has not yet
been described. Since these fixed points
may not be isolated, and the transformation may draw a
point on the stable manifold of a fixed point
$Q$ very close to another fixed point $Q'$ before
drawing it to $Q$, the difficulties of classification
and recognition of these classes
are formidable even at the conceptual level
(\cite{F2}).
Experimental uncertainties (\cite{H}) only increase them.

 Whether one is treating real systems or mathematical 
models, there are usually a number of critical
indices, some of which will be introduced explicitly
later, associated to a critical point
of the system or model. It will also
be explained, that within a
universality class, they have equal values.
The real systems are of various types: the fluids
already discussed with the liquid-gas transition;
magnetic systems, either ferromagnetic
or antiferromagnetic; mixtures of two fluids; and many 
others.
They all presumably admit an exact, although
enormously complicated mathematical description.
The best known
mathematical model of a classical physical system
is the Ising model of ferromagnetism. There are also
models, like percolation, in which a classical
thermodynamic interpretation of the parameters
is somewhat factitious.  The universality classes
cut across the classification by these features.
The principal factor is the dimension; and certain
coarse features of the local interactions, such as
isotropy or lack of it, the major secondary factor.
Other possible secondary factors are noted in \S2.6 of 
\cite{F2}.

 Our principal concern is with percolation in
dimension two; so the first factor is
fixed. Moreover there is no interaction present 
in percolation; so the second factor
is absent. The variations in lattice structure
and in percolation type, whether on sites
or bonds, that were described above appear not to
affect the universality class
of two-dimensional percolation.
 
  For systems or models to which the classical thermodynamic
paradigm is applicable, there are two quite different
types of variables: those 
that in statistical mechanics appear
as parameters in the hamiltonian
(strictly speaking, 
otherwise the temperature is not included,
in the Boltzmann weight), and in thermodynamics
are applied
externally
and naturally subject to the
control of the experimenter, the temperature and pressure 
for a fluid,
the temperature and the applied magnetic field
for a magnet; and 
those that 
it is more natural to
express 
as amounts per unit volume or
lattice site. We refer to the first as external variables
and to the second as internal.
Typical internal variables are
density, entropy
and magnetization per unit volume, or
in lattice models per site. They 
are given statistically as averages
and thermodynamically as derivatives of
a function $f$, the
{\it free energy} per unit volume or site, with respect
to a dual external variable.

 There are also two types of critical indices,
although they are not always clearly
distinguished: those associated
to thermodynamic quantities; and those that are
defined at the molecular level and
usually studied optically, or at
least electromagnetically. Although 
analogues of those of the first
type can also be defined for percolation,
the analogues of those of the second type
are the more natural in the context of this paper. The 
notion of scaling is
more easily explained for the first; so we begin with them.

 Since our treatment follows \cite {F1} and \cite{F2}, it is
more convenient to work with a ferromagnetic system.
The pertinent external variables are the temperature $T$ and
the applied magnetic field $H$. In the phase diagram 
Figure 2.2b only the curve $C$ and the point $B$ remain.
The curve $C$ is an interval, $H=0$, $T\leq T_c$, and
$B$, the Curie point, is $(T_c,0)$. The liquid-gas 
transition
is replaced by the possibility of spontaneous magnetization
along $C$ whose sign, but not magnitude, depends on whether
we approach $C$ from above or below. (Strictly speaking,
the variable $H$ is a vector, and so is the
magnetization, but this is a possibility best ignored.)

 If we choose as independent variables near $B$ 
the difference $t=T-T_c$ as well as $h=H$, so
that the critical point has coordinates $(0,0)$, then
the free energy $f=f(t,h)$ satisfies (approximately)
an equation
$$
    f(t,h)=b^{-d}f(b^{\lambda_1}t,b^{\lambda_2}h).\tag 2.2c
$$
This equation is experimental, and as explained by Fisher,
was realized by B. Widom to be a concise and
illuminating manner of expressing scaling laws.
The number $b$ is to be greater than $1$ but otherwise
arbitrary, and $\lambda_1$, $\lambda_2$ are two critical 
exponents
in terms of which all others can be expressed. For reasons
that will be discussed later, $1/\lambda_1$ is identified 
with
$\nu$. The quotient $\lambda_2/\lambda_1$ is denoted
$\Delta$. We observe that the notation for critical
indices is consistent from reference to reference, so that,
when Fisher (\cite {F1,F2}) and Grimmett (\cite{G},
especially \S 7.1 with which we urge the
reader to
compare the following
discussion) use
the same notation, they are referring to analogous 
exponents.
The integer $d$ is the dimension. Thus for the moment
it is $3$. Later, when we return to
percolation, it will be $2$.

 There are four critical indices $\alpha$,
$\beta$, $\gamma$, $\delta$ associated to thermodynamic
quantities. The induced magnetization per unit volume
is given by 
$$
  M=\partial f/\partial h.
$$
It is in essence the ferromagnetic analogue of
the density.
Taking the derivative with respect to $h$ in (2.2c),
and letting $h$ approach $0$, from above or below
for the two limits may be different, we obtain
$$
   M(t,0_{\pm})=t^{\beta}M(1,0_{\pm}),\qquad 
\beta=d\nu-\Delta,
$$
upon setting $b^{-\lambda_1}=t$. Thus near the critical 
point,
the spontaneous magnetization is (approximately!) a 
homogeneous
function of $t=T-T_c$.

 The magnetic susceptibility or
the rate of variation of $M$ with $H$, 
an analogue of the compressibility
of equation (2.2a),
is
$\partial M/\partial h$. Thus at $h=0$ it is homogeneous
of degree $-\gamma=\beta-\Delta$
as a function of $t$. The third critical index $\delta$
describes the behavior of $M$ as a function of $h$ along the
curve $T=T_c$ or $t=0$. Clearly
$$
  M(0,h)=h^{\tfrac1\delta}M(0,1),\qquad \delta=\Delta/\beta.
$$
Observe that the limit as $t\rightarrow 0$ is the same from
both sides.

 The specific heat is, apart from a factor, the second
derivative of $f$ with respect to $t$.
Thus at $h=0$ it behaves like
$t^{-\alpha}$ with $\alpha=2-d\nu$.

 There are two standard critical indices defined at the
molecular level, and therefore statistically: the 
index $\nu$ and a second index $\eta$. Away from the
critical point, correlation functions typically
decrease exponentially in space, as (very roughly)
$\exp(-\vert x-y \vert/\xi)$, where $x$ and $y$ are two 
points
in space and $\xi$ is the correlation length. At the 
critical
point $\xi$ becomes infinite and this rapid decay
is replaced by a slower decay 
$\vert x-y \vert^{2-d-\eta}$. Thus $\eta$,
in contrast to the other indices,
refers specifically to behavior 
at the critical point itself,
rather than in a neighborhood of it.

 To express $\nu$ and $\eta$ in terms of $\lambda_1$ and
$\lambda_2$ demands a more sophisticated discussion
than that for the other four critical indices (\cite{F1}).
The result is that
$$
  \nu=\frac{1}{\lambda_1},\qquad \eta=2-\frac{\gamma}{\nu}=
   2+d-2\lambda_2
$$
As a consequence, $\lambda_1$, $\lambda_2$, and all the 
other
critical indices can be expressed in terms of $\nu$ and 
$\eta$.

 Scaling is a statement about a specific physical system
or model. Universality, which asserts that the critical
indices are constant (or nearly so) on broad classes
is a second, quite distinct assertion. The evidence
for both consists largely either of experimental data
or the results of computations for specific models.

 Theoretical justification is scant. The renormalization
group yields, however, some  insight into (2.2c). It is 
easiest
to consider lattice models of ferromagnetism, in which 
each site of 
the lattice $L\subset {\Bbb R}^d$ of \S 2.1 is taken to be 
occupied by a 
magnet, whose magnetization and orientation may
or may not be sharply
constrained. 
In the widely studied Ising model it is constrained
to take either of two opposing orientations
and to be of fixed magnitude,
thus effectively to assume only the values $\pm1$.
Constraints are unimportant
at the moment; it is rather the geometry that counts.
Rather than taking  only simple magnets at the sites, we 
could
also allow some complicated system formed by
a collection of mutually interacting
magnets to be the object attached to the site. Then the 
interaction
between the objects at neighboring sites,
or more generally sites in close proximity, will be the
resultant of the interaction between the magnets in the
systems attached to the two sites. The 
advantage of the more general formulation is
that such systems can be composed.

 This is the essence of renormalization, and the 
expository problem at this 
point is to provide the reader with
some idea of this composition, because
it informs all our investigations,
but without prejudicing in any way
the precise form it is to take.
It is not the least of our purposes (as in \cite{L1,L2})
to search for novel, perhaps even mathematically more
tractable definitions of the composition. 

 We begin vaguely. The systems
attached to the sites at the corners of a 
$d$-dimensional cube can be fused into a single
system. Starting therefore with one model $M$, we can
construct a second $M'=\Theta(M)$
by attaching to the site 
$x=(x_1,x_2,\dots,x_d)\in \Bbb R^d$ the system
obtained by fusion from those at the
sites
$$
   x'=(2x_1+\epsilon_1,2x_2+\epsilon_2,\dots,2x_d+
\epsilon_d),
$$
the numbers $\epsilon_i$ each taking the values $0$
and $1$. 

 Consider, as in \S 2.1 the system formed by
the magnets on the 
sites inside a large block $S_n$ of side $n$. If $n=2^m$
the system is obtained by starting with independent systems
of side $1$, putting $2^d$ together to form a block of side
$2$, and then iterating the procedure
$m$ times. Thus the model $M$ in the bulk can be
considered to be the model $M^{(m)}=\Theta^m(M)$.
Since the basic assumption of statistical mechanics
is that the properties of sufficiently large finite systems
are essentially those of infinite systems, we might
suppose that $M^{(m)}$ and $M^{(m+1)}$ were
essentially the same; thus, that $M^{(m)}$ was a fixed 
point of
$\Theta$.

 The mapping $\Theta$ is a {\it renormalization}, so that 
fixed points of the (semi-)group it generates appear
to be objects of central importance.
Universality can now be formulated
as the assertion that there are
few fixed
points of $\Theta$ pertinent to the
systems of interest.

 The first, obvious
difficulty is that to define $\Theta$ we have had to
allow our system to grow more complex, so
that a problem of closure presents itself.
The second, less obvious, is that although
what may be one of the major factors
responsible for universality
is implicit in
the definition of $\Theta$, nothing
in the definition provides any insight into
the mechanism by which it prevails over
the details of the local interaction. Namely,
the propagation in
$M'=\Theta(M)$ is across the walls separating the
$2^d$ constituents of the composite system, and as we
iterate $\Theta$ the
number and nature of the paths along
which the system at one site influences
those at another depend strongly on the dimension
$d$, and this multiplicity appears to dominate
all other factors.

 In one dimension the propagation is linear,
and the problems
can usually be formulated in terms of Markov
processes, so that 
an analysis in terms of the
renormalization group, although 
instructive,
is from a strictly mathematical
point of view not necessary.
In two and more dimensions,
it is one of the most effective
methods for obtaining a handle
on the qualitative behavior of the system
at a critical point, but the problem of closure
becomes more severe (\cite{F2,\S5.6}).

 Although the crossing probabilities of
the next section are the coordinates
whose utility in the study of
renormalization we are examining,
standard treatments 
more often use, in one form
or another, the external variables
that appear in the hamiltonian. A simple
example due to  Nelson and Fisher and taken
from \S 5.2 of \cite{F2} admits a precise
definition of renormalization, and may give the 
reader a clearer notion of the way it functions.

 It is the Ising model in one dimension. Consider a
finite collection of integers $S_N=\{i\vert -N\leq i\leq 
N\}$.
The possible states of the model are 
the functions $s$ on $S_N$ with values in $\{\pm 1\}$.
The energy of a state is given by the hamiltonian
function,
$$
  H_0(s)=K_0\sum_{-N\leq i\leq N-1}s_is_{i+1}+
          h_0\sum_{-N\leq i\leq N}s_i+C_0\sum_{-N\leq 
i\leq N}1.
$$
In statistical mechanics the free energy per site is
given as the quotient
$$
 -kT\ \ln(\sum_s\,\exp(-\frac{1}{kT}H_0(s)))/(2N+1)
$$
the sum running over all states $s$. 
(The factor $k$ that ensures that the argument
of the exponential function is dimensionless
is called the Boltzmann constant.) Emphasis is 
therefore often
put on the {\it partition function}
$$
  Z_N(H)=\sum_s\,\exp(-\frac{1}{kT}H_0(s))=\sum_s\,%
\exp(-H(s)),\tag 2.2d
$$
where we have set
$$
  H(s)=H(s;K,h,C)=K\sum_{-N\leq i\leq N-1}s_is_{i+1}+
          h\sum_{-N\leq i\leq N}s_i+C\sum_{-N\leq i\leq N}1,
$$ 
with
$$
     K=\frac{K_0}{kT},\qquad h=\frac{h_0}{kT},\qquad 
C=\frac{C_0}{kT}.
$$
It is appropriate to refer to $K$, $h$, and $C$ as the 
external
variables. (There is, as observed, a slight 
abuse of terminology here. The parameter
$T$ appears in $K$ but not, strictly speaking, in the
original hamiltonian.) Observe that in statistical mechanics
the probability of the state $s$ is taken to be
equal to
$$
   \exp(-H(s))/Z_N(H).
$$

 We could fuse the systems at $s_{2i}$ and $s_{2i+1}$
so that the system attached to the site $i$
then consisted of two simple magnets interacting
through the energy $K's_{2i}s_{2i+1}$, but this changes
the nature of the system, so that problems
of closure arise. Rather the emphasis is put on calculating
the partition function as a function of the three 
external parameters.  Fix the values of the $s_{2i}$ so that
the local state is determined at the even sites, and
take the sum in (2.2d) over the two possible values of 
$s_{2i+1}$
at all the odd sites. If we define $s'$ by $s'_i=s_{2i}$, 
the result
may be written as
$$
  \sum_{s'}\exp(-H'(s'))=Z'_{N'},\qquad N'=N/2,
$$
if a certain fuzziness at the endpoints is accepted.
It can be expected to resolve itself in the limit
of large $N$. The problem of closure arises because
the hamiltonian $H'$ may be of quite a different form
than $H$, so that the calculation transfers us to a
larger space of hamiltonians, and no real simplification
has been achieved. 

 The advantage of the example
(we stress that it is very unusual),
achieved only at the cost of
abandoning the initial fusion
and 
summing in an arbitrary manner
over the 
states at the odd-numbered sites, is that $H'$ turns
out to be of the form 
$$
H'(s')=H(s';K',h',C')\tag 2.2e
$$
if 
$$
\aligned
w'&=w^2 x y^2/ (1+y)^2(x+y)(1+xy)\\
x'&=x(1+y)^2/(x+y)(1+xy)\\
y'&=y(x+y)/(1+xy).
\endaligned 
$$
The three parameters appearing here are given by
$$
 w=e^{4C}, \qquad x=e^{4K}\quad \text{and}\quad y=e^{2h}.
$$
Thus  $\Theta$ appears here simply as the transformation
$$
 (K,h,C)\rightarrow (K',h',C').
$$

 In order to examine the physical properties of the 
hamiltonians $H$, one
can use the correlation length $\xi(H)$. Let $f(i)$ be the
(limit for large $N$ of the) probability
that $s_0$ and $s_i$ have the same orientation and let 
$\xi (H)$
be a measure of the width of this distribution, say the 
largest value $\vert
i\vert$ such that $f(i)>f_0$ for some constant $f_0$. 
If we limit ourselves to 
the even integers $i$,
the value of $\xi$ should not
change seriously, so that
partial summation over the
odd sites does not affect
the correlation length.
On passing from $H$ to $H'$ we relabeled, denoting
$s_{2i}$ be $s'_i$.
The result is therefore that
$$\xi(H')= {1\over 2} \xi(H).$$
The \gr transformation $\Theta$ is, in this example,
the process of ``decimation'',
thus of removing one-half the sites, followed
by a shrinking of the lattice scale, and the replacement 
of $H$ by $H'$.
It decreases the correlation length by the factor $1/2$. 
The space of models
is parametrized by 
a subset of $\Bbb R^3$. 

 We have claimed that the fixed points of
the map $\Theta$ are of major interest. At a fixed point 
$(w,x,y)$ of the map $\Theta$ the hamiltonian
$H=H(w,x,y)$ would be invariant under $\Theta$ and its
correlation length would have to satisfy
$$
\xi(H)={1\over 2}\xi(H),
$$
so that $\xi(H)=0$ or $\xi(H)=\infty$. For the physical
reasons explained at the beginning of this section,
it is the solutions of the second type that
yield critical points. 
They are examined in more detail
in \S 5.3.2 of \cite{F2} and in \cite{NF}.

 The simplicity of this one-dimensional example is 
misleading. 
For the two-dimen\-sional Ising model, 
decimation appears to require the
introduction of a further variable (in addition to $K, h$ 
and $C$)
describing the interaction of second nearest neighbors.
Iterating the decimation
will require more and more variables, so that the 
problem of closure manifests
itself clearly.
This behavior, and not that of the example, is typical.
What one expects in general is that
(2.2e) will be replaced
by an equation
$$
H'(s')= H(s', K', h', C') + H''(s'),\tag2.2f
$$
in which $H''(s')$ is small, and at each step
smaller, eventually becoming {\it irrelevant}.

 In \cite{L1} and \cite{L2}
the emphasis is on approximations
to the ``true'' $\Theta$ by a collection of increasingly
complex transformations
that act on finite-dimensional spaces and whose first
members permit close study. Since 
these approximating transformations are the
reason for our
emphasis on crossing probabilities, we shall
briefly describe them in the next section.

 We first return briefly to the equation
(2.2c) imagining ourselves at a fixed point.
It will be associated to a complicated system, so
that there will be many more external variables
than merely $h$ and $T$ (or $t$) needed to determine
the local interactions and therefore the free energy 
per site but, typically, they will be irrelevant.
Mathematically this means that they are variables along the
directions in which $\Theta$ is contracting.
(In the example, $K$ is just another form of $T$, but as 
often
happens, there is more than one supplementary relevant
variable, not only $h$ but in addition $C$. The irrelevant
variables, had they appeared, would be those defining 
$H''$.)
If we ignore these irrelevant directions then
$\Theta$ will be roughly of the form
$$
  (t,h)\rightarrow (2^{\lambda_1}t,2^{\lambda_2}h).
$$
Since renormalization obviously multiplies the free energy 
per
site by $2^d$, we obtain, upon
ignoring the other, irrelevant variables,
the equation
$$
   2^df(t,h)=f(2^{\lambda_1}t,2^{\lambda_2}h).
$$
Iterating we obtain (2.2c) with $b$ equal to a power
of $2$. In other words, scaling can be recovered
from renormalization group arguments. So can universality,
because the two indices $\lambda_1$ and $\lambda_2$
are associated to the fixed point, not to the model
with which the iteration begins. 

 It is implicit in these equations that for
$$\
   \vert t\vert  +\vert h\vert=1
$$
the value of $f$ is neither very large nor
very small.
Both $\lambda_1$ and $\lambda_2$
are positive. Since $f$ is the free energy per site,
it is clear 
from (2.2c) that for $t$ and $h$ very small,
the side, $b$, of the block needed in order that the
total free energy be of order $1$ is given
by the condition
that
$b^df(t,h)\sim1$, thus that $b=t^{-\lambda_1}$
or $b=h^{-\lambda_2}$. If $h\ll t$, the first
condition gives the smaller $b$ and the relation
$\nu=1/\lambda_1$. For more serious demonstrations of
this relation the reader is referred to \cite{F1}.

 We observe in passing that
$\lambda_2$ can be larger than 
$\lambda_1$ so that we see no very strong reason that
$$
  f(t,h)=b^{-d}f(b^{\lambda_1}(t+ch),b^{\lambda_2}h),
$$
with a constant $c$
might not be preferable to (2.2c). We have followed 
convention.

\subheading{\num{2.3.} Crossing probabilities}
 Percolation is not a model
of a classical physical system with
a thermodynamic interpretation, and the finite
models that appear 
later in this section are
stripped of many features
of such models; so their value is uncertain.
Their purpose,
as we have already remarked, is to provide a model of the 
dynamics
of renormalization that is accessible mathematically,
and that reveals the essence of the processes involved.
It is still far from  certain that this purpose
will be achieved, but to defend it as a goal
we cite a phrase from Fisher's description
in \cite{F1,\S 1.2} of the role of models:
{\medskip\narrower\narrower\noindent
``\dots the aim of the theory of a complex phenomenon 
should be
to elucidate
which general features \dots of the system
lead to the most
characteristic and typical observed properties.''
\smallskip}
\noindent
We have deleted the words ``of the Hamiltonian'' because
we focus on percolation, deliberately to avoid all problems
caused by the hamiltonian. Those caused by the multiple
paths along which effects are propagated in two dimensions 
remain,
so that Fisher's demand that initially:
{\medskip\narrower\narrower\noindent
``\dots one should aim at a broad qualitative 
understanding, successively
refining one's quantitative grasp of the problem''
\smallskip}
\noindent
is met.

 The rest of the paper concentrates on two-dimensional 
percolation. 
The two hypotheses presented in \S2.4
relate the critical behavior of a large class of models. 
Before stating
these hypotheses we shall first introduce the models they 
are likely to 
describe and then extend the notion of
the horizontal crossing probability $\pi_h$ to
larger families of geometrical data.

Let $\Cal G$ be a graph embedded in $\Bbb R^2$. As in the 
introduction, we
refer to its vertices as sites and to its edges as bonds. 
It is a {\it
periodic graph} \cite{K} if it satisfies the following 
conditions:
\roster
\item $\Cal G$ contains no loops (in the graph-theoretical 
sense);
\item $\Cal G$ is periodic with respect
to translations by the elements of a lattice 
$L$ in $\Bbb R^2$ of rank two;
\item the number of bonds attached to a site in $\Cal G$ 
is bounded;
\item all bonds of $\Cal G$ have finite length and every 
compact
set of $\Bbb R^2$ intersects finitely many bonds of $\Cal 
G$;
\item $\Cal G$ is connected.
\endroster
Let $\Cal S$ be the set of sites of $\Cal G$ and $p:\Cal S 
\rightarrow
[0,1]$ a periodic function, 
thus a function invariant under the translations
from $L$. As before we allow each site $s\in \Cal S$ to
be in either state $0$ (closed) or $1$ (open) and we 
define a measure  $P_s$ on
the set $\{0,1\}$ by 
the equations $P_s(0)=1-p(s)$
and $P_s(1)=p(s)$.
Finally we introduce the set of configurations
$X$ on the graph $\Cal G$ as the product $\prod_{\Cal S} 
\{0,1\}$ and endow
$X$ with the product measure $m$ of the various $P_s$.
A model $M=M(\Cal G, p)$ is defined as the set of data 
$\{\Cal G, p, X, m\}$.
We shall refer to these models as the class of  {\it 
graph-based models}.
Observe that for a given $\Cal G$ the family of possible 
functions
$p$ form a compact set in some finite-dimensional space.

 The model $M_0$ corresponds to a graph constructed of the 
vertices $\Bbb Z^2$
with edges between nearest neighbors and the function $p$ 
constant on all
sites. The definition 
also includes the models of percolation by
sites on triangular and hexagonal lattices. 
To include models of percolation by bonds 
one associates (\cite{K}) to a graph
$\Cal G$ its matching graph $\tilde\Cal G$. The sites of 
$\tilde\Cal G$ are the
midpoints of the bonds of $\Cal G$; two distinct
sites $\tilde s_1$ and $\tilde s_2$
of $\tilde\Cal G$ are joined if and only
if the corresponding
bonds $b_1$ and $b_2$ of $\Cal G$ are attached to a common 
site.
A periodic function $p$ on the {\it bonds} of $\Cal G$ 
leads naturally to
a periodic function $\tilde p$ on the {\it sites} of 
$\tilde\Cal G$ and we
can therefore replace percolation by bonds on $\Cal G$ by 
percolation by
sites on $\tilde\Cal G$. Percolation by bonds on a square
lattice where horizontal bonds are open with probability 
$p_h$ and vertical
ones with
a different probability $p_v$ is an example of 
a model for which the probability
function is not constant.

 The hypothesis of universality in \S 2.4 
has only been examined
numerically for a few models.
If we were eager to be precise, we
might suggest the
class of graph-based models
as the appropriate class
for which to\ formulate
the hypothesis.
Such precision is inappropriate
at this stage.
In particular, other models
will very likely fall into the same 
universality class.

 That this is so
for a model based on 
an aperiodic graph whose sites and bonds are defined
by a Penrose tiling on the plane is indicated
by the results of \cite{Y}. 
Thus the condition of periodicity is excessively prudent.
Models may also be defined 
without any reference to graphs, for example
by  randomly placing
unit disks on the plane $\Bbb R^2$ 
with a density $\delta$.  If a 
rectangle is drawn on the plane, a horizontal crossing  
is a path from left to right on overlapping disks. The 
density
$\delta$ plays the role of the probability $p$ 
that a site is open.
(See \cite{G} for a discussion of the ``snails on a lily 
pond'' model.)
The disks 
can be replaced by ellipses with uniform random 
orientation or, in the limit, by segments of length one.
Results of H.~Maennel
for crossing probabilities 
in this limiting case confirm
that 
they are the same as those of $M_0$.

 For graph-based models the notion of a cluster
for a given state is
simple. It is a maximal connected subset
of the set of open sites.
The universality emphasized in \cite{U} is that of the
crossing probabilities, the probabilities 
of events defined by a
simple closed curve $C$ in the plane and by arcs
$\alpha_1,\dots,\alpha_m$, and $\beta_1,\dots,\beta_m$, as 
well as
$\gamma_1,\dots,\gamma_n$ and
$\delta_1,\dots,\delta_n$ of $C$.

 Let $A$ be a large constant and define $C'$ and the 
intervals
$\alpha_i'$, $\beta_i'$, $\gamma_j'$, and $\delta_j'$ to 
be the
dilations, with respect to some
fixed but irrelevant point in the plane,
of $C$ and $\alpha_i$, $\beta_i$, $\gamma_j$ and 
$\delta_j$ by
the factor $A$. In principle
a given state admits a crossing
inside $C'$ from $\alpha_i'$ to $\beta_i'$ if there is 
cluster for this state whose intersection with the 
interior of
$C'$ intersects both $\alpha_i'$ and $\beta_i'$. Since 
$C'$ is a curve,
it might not contain any sites and it is in fact
necessary to replace $C'$, supposed to be not too irregular,
by a band, and to thicken the intervals accordingly. 
Then there will be a crossing between $\alpha_i'$ to 
$\beta_i'$ if
there is an open path inside $C'$ from the thickening of 
these two
intervals. For large
$A$ the choice of band, provided it is relatively narrow, 
is irrelevant.
We describe specific conventions when discussing the 
experiments.

\fighere{16.5 pc}\caption{Figure 2.3\rm a. Data 
$(C,\alpha,\beta,\gamma,
\delta)$ defining the event $E$.}
\eject

 With appropriately chosen conventions we can therefore 
define
$$
\pi_A(C,\alpha_1,\dots,\alpha_m,\beta_1,\dots,\beta_m,%
\gamma_1,\dots,\gamma_n,
        \delta_1,\dots,\delta_n),
$$
the probability that there are crossings in the interior 
of $C'$ from
$\alpha_i'$ to $\beta_i'$ for $1\leq i\leq m$ but no 
crossing
from $\gamma_j'$ to $\delta_j'$ for $1\leq j  \leq n$. One 
may suppose
that these conventions
will be such that they do not affect the existence
or the value of the limit
$$
    \lim_{A\to\infty}
   \pi_A(C,\alpha_1,\dots,\alpha_m,\beta_1,\dots,\beta_m,%
\gamma_1,\dots,\gamma_n,
        \delta_1,\dots,\delta_n)=
      \pi(E,M). \tag 2.3a
$$
We take $E$ as an appropriate abbreviation for the event (or
rather events since we took a limit over dilations)
defined by $C$, $\alpha_i$, $\beta_i$,
$\gamma_j$ and $\delta_j$. The horizontal crossing 
probability $\pi_h$ 
defined for $M_0$ in the introduction is
a special case of $\pi(E, M_0)$. The curve $C$ is a square 
and
only two arcs $\alpha$ and $\beta$ are chosen, the left 
and right sides.

 A natural extension (\cite {K}, \cite{AB}) of the theorem 
 of 
\S 2.1 is that, a family
of models $M(\Cal G, p)$,
parametrized by the function $p$,
is constituted by two 
open sets, one for
which the limit (2.3a) is always $1$ and one for which it 
is always
$0$; a third subset, the set of critical probabilities, 
separates the other 
two and is such that the limit (2.3a) (if it exists) lies 
in general between
$0$ and $1$. Presumably the limit does exist even for the 
critical probabilities,
but this has not yet been established. The two simplest 
models,
percolation by sites and bonds on a square lattice,
for which $p$ varies over an interval,
are critical
for a single appropriate choice $p_c$ of $p$. Hence the 
two open sets are
$[0,p_c)$ and $(p_c,1]$ and the critical subset is 
$\{p_c\}$.
For percolation by sites
the value of $p_c$ is known empirically to be 
$0.5927460\pm 0.0000005$ \cite{Z};
for percolation
by bonds it is known theoretically \cite{K} to be 
$\tfrac12$.

 All our numerical work, as well as the hypotheses 
underlying it,
is predicated on the existence of these limits, that we 
now take 
for granted. Moreover our models are from now on supposed 
to be critical.

 Since our investigations
were initially prompted
by the desire to
provide empirical foundations
for the definitions
of the finite models
of \cite{L1} and \cite {L2},
we review those definitions
briefly. We shall also
need to have them at our disposal
in \S 2.5.

 Let $S$ be a square whose sides have been divided in $l$ 
equal intervals.
There are $4l(4l+1)/2$ pairs of intervals. Let $\Cal P$ be 
the set of these
pairs. A configuration $x$ for this model is 
obtained by 
specifying which pairs are
connected and which ones are not. 
Assign them respectively the values $+1$ and $-1$.
The space $A$ of configurations
is then a set of functions from $\Cal P$ to $\{+1, -1\}$.
(There are technical constraints
on the configurations that need not be described here.) 
Therefore
each element of $A$ is an 
event $E$ whose defining curve $C$ is a square.
(According to the hypothesis of conformal invariance,
all crossing probabilities can be obtained
from those for this case. See \S 2.4)

 There is a natural transformation 
$\Theta_A:A\times A\times A\times A\rightarrow
A$ that is similar to the \gr transformation $\Theta$ of \S
2.2. To construct $\Theta_A$ one first juxtaposes 
four elements of $A$ so that they\ form
a larger square with $2l$ subdivisions on its sides. These
intervals 
are then fused in pairs so that each side of the larger 
square 
contains $l$ 
intervals. Finally these new intervals are
connected by composing the ``paths''. Suppose, for 
example, that $\alpha$
and $\beta$ are connected intervals in one of the original 
squares and
$\mu$ and $\nu$ are also connected in another one. If 
$\beta$ and $\mu$
turn out to be in
the interior of the larger square 
formed upon juxtaposition and are coincident,
then the larger intervals containing $\alpha$ and $\nu$ in 
this
square will be connected. See Figure 2.3b for an example.

\fighere{11.5 pc}\caption{Figure 2.3\rm b. An example of 
the 
transformation $\Theta_A\: A\times A\times A\times A\to A$ 
for a finite model.}

If $\frak X$ is the set of 
measures on $A$, $\Theta_A$ can be used to define a map
$\Theta_{\frak X}:{\frak X}\rightarrow {\frak X}$. Since 
$\frak X$ is a simplex in a finite-dimensional
space, the question of finding fixed points of 
$\Theta_{\frak X}$ and studying
their nature is well-posed. 

\subheading{\num{2.4.} The two hypotheses}
Although Aizenman prefers to distinguish between the 
hypothesis
of universality and that of conformal invariance, 
regarding the first
as commonly accepted, even in the form in which we state it,
we prefer for the sake of clarity as well as for the reasons
already rehearsed in \cite{U} to state them in a 
less invidious form. The purpose of \cite {U} was to show 
that
the probabilities $\pi(E,M)$ were independent of $M$, 
provided
the model satisfied some simple conditions of symmetry. 
This is a form
of universality. To state the general form we observe that
the group $GL(2,\Bbb R)$ acts independently on the models
and on the events. (From now on we restrict ourselves to 
the 
class of graph-based models.)

 A model with sites $\{s\}$ and bonds $\{b\}$ is sent by 
$g\in GL(2,\Bbb R)$ to the model with sites $\{gs\}$ and 
bonds
$\{gb\}$, the probability function $p$ being transferred
directly from the old sites and bonds
to the new. The lattice $L$ defining the periodicity is
then replaced by $gL$. 
The group $GL(2,\Bbb R)$ acts on the events
$E$ as well. We shall write $gE$ for the event obtained from
the data $(C, \alpha_i, \beta_i, \gamma_j, \delta_j)$ 
defining $E$
by letting $g$ 
act on each element of the data:
$gE=(gC, g\alpha_i, g\beta_i, g\gamma_j,g\delta_j)$.
By the definitions,
$$
\pi(gE,gM)=\pi(E,M),$$
since transforming simultaneously the embedding of the 
graph $\Cal G$ and
the curve $C$ by the same linear transformation does not 
alter
$\pi(E,M)$.
On the other hand, the probabilities
$\pi(E,gM)$ and $\pi(gE,M)$ are generally quite
different from $\pi(E,M)$. 

\proclaim {Hypothesis of Universality} If $M$ and $M'$ are 
any two (graph-based)
models
of percolation there is an element $g$ in $GL(2,\Bbb R)$
such that 
$$
   \pi(E, M')=\pi(E,gM)  \tag 2.4a
$$
for all events $E$.
\endproclaim

 Those experienced readers who feel that this hypothesis 
is generally accepted,
and not worth examining numerically,
might ask themselves how much they are willing to stake on 
its validity
in three dimensions --- life, family, career?
Less experienced readers will be more likely
to notice just how strong
the statement is, and therefore
to be more skeptical. We ourselves
have found an explicit
enunciation a great aid to clear thinking. 

 In paragraph 3.4, we shall give an example of a
model $M$   
for which the matrix $g$ of the hypothesis has no
elements equal to $0$.

 The hypothesis obtains its full force
only in conjunction with that of conformal
invariance. Suppose that $J$ is a linear transformation of 
the plane $\Bbb R^2$
with $J^2=-I$. Then $J$ defines a complex structure on the 
plane,
multiplication by $i$ being given by
$x\rightarrow Jx$. Once $J$ is fixed, the notion of a 
$J$-holomorphic
map on an open subset of the plane can be introduced as 
well as that
of an antiholomorphic map. 
If $g\in GL(2,\Bbb R)$ and $J'=gJg^{-1}$, then the map 
$\phi\rightarrow g\cdot \phi\cdot g^{-1}$ transforms 
$J$-holomorphic
maps into $J'$-holomorphic maps and $J$-antiholomorphic
maps into $J'$-antiholomorphic maps.

 If $\phi$ is a transformation $J$-holomorphic
in the interior of $C$ and continuous 
and bijective up to its boundary, which is just $C$ itself,
then the event $\phi E$ is well defined; the 
transformation $\phi$
is simply applied to the data $(C, \alpha_i, \beta_i, 
\delta_j, \gamma_j)$
defining $E$. We may also apply a 
transformation $\phi$ that is antiholomorphic in the 
interior to 
$E$. The following hypothesis was in essence suggested
by Michael Aizenman.
\proclaim{Hypothesis of conformal invariance} For every 
model 
$M$ there is a linear transformation $J=J(M)$ defining a 
complex
structure such that
$$
  \pi(\phi E,M)=\pi(E,M) \tag 2.4b
$$
for all events $E$
whenever $\phi$ is $J$-holomorphic or $J$-antiholomorphic
in the interior of $C$ and continuous (and,
for the moment, bijective) up to its boundary.
\endproclaim
 To understand the nature of the hypothesis,
consider the model $M_0$ of percolation by sites
on a square lattice. The complex
structure for $M_0$ is, if the hypothesis is correct,
the usual one defined by 
$$
J_0=\pmatrix 0& -1\\1& 0
\endpmatrix
$$
and the associated holomorphic functions are the usual
ones.

 Given an event $E$ we may choose
$\phi$ so that $E'=\phi E$ is defined by 
the the unit circle $C'$
with centre at the origin and arcs on it.
If for example
$E$ is defined by the horizontal
crossing of a rectangle, then the data on
$C'$ will be four points
$a$, $b$, $c$, and $d$, the images of
the four corners of the square under $\phi$,
and $\alpha'=\phi(\alpha)$ will be the 
circular arc between $a$ and $b$,
and $\beta'$ the circular arc 
between $c$ and $d$.

 For numerical work it is easier
to use the inverse of $\phi$, a  Schwarz-Christoffel 
transformation
$$
\psi:w\rightarrow\int_0^w {du\over \sqrt{(u^2-v^2)(u^2-1)}},
$$
in which $v$ is a constant
of absolute value $1$ that depends on the aspect
ratio of the rectangle. For a square, one can clearly take
$v=\sqrt{-1}$. In the arguments of \S 2.6 the disk is
replaced by the upper half-plane, and in \S 3.5 the
hypothesis is implicitly reformulated for
all unbounded regions. 

 If $M$ and $M'$ are related by the first hypothesis then
$J(M')=gJ(M)g^{-1}$. Denote the identity
transformation by $I$. The set of linear transformations

$$
   H(J)=\{aI+bJ\in GL(2,\Bbb R)\vert a^2+b^2\neq0\}
$$
is the centralizer of $J$ in $GL(2,\Bbb R)$, and is of 
index two in
$$
    H'(J)=\{h\in GL(2,\Bbb R)\vert hJh^{-1}=\pm J\}.
$$
The group $H$ determines $H'$ but only determines $J$ 
itself up to sign.
If $J=J(M)$ we write $H(J)=H(M)$ and $H'(J)=H'(M)$. It is 
clear that the
element $g$ that appears in the hypothesis of universality 
is
not uniquely determined, at best the class $gH'(M)$ is 
determined.
As we shall observe explicitly later, there is in fact no 
further
ambiguity, so that the two hypotheses together imply
that the image
under
$$
   \psi: M\rightarrow\prod_E \pi(E,M)\tag 2.4c
$$
of the set of all models
in the product, over all events $E$, of the interval $[0,1]$
(a very, very large set) is 
a small subset that may be identified with the
upper half-plane. Each model of the class defined
in \S 2.3 corresponds to point in the upper-half plane.
All the crossing probabilities $\pi(E)$ of models 
corresponding to the
same point are identical. Thus universality
and the orthogonal invariance
of $M_0$ reduce
an apparently infinite-dimensional continuum of 
possibilities
for the image of $\psi$ to a two-dimensional
continuum. Without orthogonal invariance, this continuum
would already be three-dimensional; so universality is
the determining factor. 

 Those who have read \S 2.2 will notice that the
universality of that section is quite different from
that of this paragraph. Universality\ in \S 2.2 is that of
{\it critical exponents}
and they could all be expressed in terms of $\lambda_1$
and $\lambda_2$ that can themselves be interpreted as
the logarithms of 
the dominant eigenvalues of the
Jacobian matrix of a suitable
{\it renormalization} transformation
at a fixed point. This fixed point is not
usually regarded as existing
in a physical sense,
and 
is therefore
treated as a somewhat spectral object.
The assumption implicit in the finite
models mentioned in \S 2.3 is that the fixed point 
itself, at least for percolation, is a real physical
and mathematical
object whose coordinates are the {\it crossing
probabilities}, so that not only the critical indices
but also these probabilities are universal. They and not
the critical indices are the
objects of principal interest in this paper.
Nevertheless, although -- mathematically -- the point
and its coordinates have to be studied before 
the eigenvalues of a transformation fixing it,
it is the critical indices
whose universality is to be explained
and that have attracted the
most attention 
from physicists so far. It is by no means certain
that for other problems than percolation there will
be useful analogues of the crossing probabilities
of \S 2.3, and even less clear that they will
be physically significant.
 
 Although we do not want the renormalization
group to intrude
too obstreperously
on the discussion, we repeat, in order that there be no
misunderstanding, that the crossing probabilities
are not to be interpreted as coordinates
of the model at a critical
value of the parameters but as
those of the fixed point to which it is attracted.
This is what permits the image of the map $\psi$
to be of such a small dimension.

 To be concrete the image (2.4c) is obtained as the 
collection
$$
   \psi: M\rightarrow\prod \pi(E,g^{-1}M_0),\qquad g\in 
GL(2,\Bbb R),
$$
where $M_0$ is a given model, and the half-plane is
identified with
$$H(M_0)\backslash GL(2,\Bbb R).$$
Observe that the action of $GL(2,\Bbb R)$
on this homogeneous space
is to the right and is given on coordinates
by
$$
      \pi(E,M)\rightarrow\pi(gE,M).
$$
The image (2.4c) can be identified with the set of all 
possible groups $H(M)$, thus with the set of all 
translation-invariant
conformal structures on the plane up to orientation.

 In a certain sense the hypothesis of universality is 
subsumed under
that of conformal invariance, because the relation (2.4a) 
may be written
$$
\pi(E, M')=\pi(g^{-1}E,M),
$$
and $g^{-1}$ is a translation-invariant conformal map from 
the
structure defined by $H(M')$ to that defined by $H(M)$, 
thus,
in general,
between two {\it different} conformal structures. 
The two hypotheses are thus fused into one if the
equation  
$$
\pi(E, M')=\pi(\phi E,M)
$$
is supposed valid for any map $\phi$ that is defined on the 
interior of the curve $C$ determining $E$ and
continuous up to its boundary, and
takes the conformal structure attached to $M'$ to
that attached to $M$.

 Since $H'(M_0)$ contains  
the reflections in both axes as well
as the permutation of the two axes, it
must be the orthogonal group,
and we
can identify the image (2.4c) with
the upper half-plane in such a way that
$M_0$ corresponds to the point $i$.
The action of $GL(2,\Bbb R)$ is then
$$
   \left( \matrix a&b\\
               c&d\endmatrix\right):z\rightarrow \frac{az+
c}{bz+d},
$$
when $ad-bc$ is positive, and is
$$
   \left( \matrix a&b\\
               c&d\endmatrix\right):z\rightarrow 
\frac{a\bar z+c}{b\bar z+d}
$$
otherwise.
Let $R$ be the group of four matrices
$$
  \left(\matrix \pm1&0\\
              0&\pm1\endmatrix\right)
$$
and $S$ the group generated by $R$ and the
matrix
$$
   \left(\matrix 0&1\\
              1&0\endmatrix\right).
$$
A simple calculation shows that the points invariant under 
$R$ are the points on the imaginary axis, and that the only 
point invariant under $S$ is the point $i$ itself. 

 In \cite {U} we studied only models that obviously 
yielded points
invariant under $R$, and thus were implicitly confining
ourselves to a one-dimensional curve, the imaginary axis,
in an otherwise two-dimensional family.

\subheading{\num{2.5.} More critical indices for 
percolation}
 As we saw in \S 2.2 it is natural in models
and systems with a thermodynamic significance
to emphasize the 
way in which the internal variables
depend on the external ones, and thus to introduce the 
critical indices
$\alpha$, $\beta$, $\gamma$ and $\delta$.
Once we pass to other coordinates,
or other models in which there is no {\it natural}
choice of coordinates, it is
no longer clear which are the principal
critical indices.

 The abstract possibility of blowing up or contracting
the ill-defined space in which $\Theta$ operates
creates even more ambiguity.
Suppose, for example, that in some rough sense
$\Theta$ operates in the neighborhood of a fixed point
as
$$
  \Theta:\quad (t_1,t_2,t_3,\dots)\rightarrow
   (2^{\lambda_1}t_1,2^{\lambda_2}t_2,2^{\lambda_3}t_3,%
\dots),
$$
and that only $\lambda_1$ and $\lambda_2$ are positive,
so that only the first two coordinates
are relevant.
If we allow ourselves that freedom,  then blowing up,
as usual in algebraic geometry, so that
$(t_1,t_2/t_1,t_3,\dots)$ or
$(t_1/t_2,t_2,t_3,\dots)$
become the coordinates, we replace $\lambda_2$
by $\lambda_2-\lambda_1$ or
$\lambda_1$ by $\lambda_1-\lambda_2$, creating two fixed
points from one, and perhaps changing the number of
unstable variables.

 For percolation itself, our preferred coordinates
are the numbers $\pi(E,M)$ defined by crossing
probabilities. These permit readily, as we saw in \S 2.1,
the introduction of the critical index $\nu$.
Although the critical indices $\alpha$, $\beta$,
$\gamma$, and $\delta$
can be defined directly within percolation (\cite{G}),
that they are indeed the analogues of those
of \S 2.2 is best seen as in
\cite {E2,\S2} by treating percolation
as the limit of an Ising model in a weak field.
They do not have an obvious interpretation
in terms of the crossing probabilities that are in this 
paper
the primary objects.

 This can perhaps be forgiven if we
can at least interpret $\eta$, which we recall
refers to behavior at criticality, in terms 
of crossing probabilities. To this
end we borrow some
standard conjectures from \cite{G, Chap.~7}, and use
freely the notions of conformal invariance developed
in Part Three. We work with the model $M_0$ at $p=p_c$.

 Let $P(r)$ be the probability
at $p=p_c$ that the origin is open
and the cluster containing it also contains
a point at a distance at least $r$ from the origin. It is
believed \cite{G, (7.10,7.11)} that
$$
    P(r)\sim r^{-1/\rho},\qquad \rho=48/5.\tag 2.5a
$$
If $z$ is a point in the lattice ${\Bbb Z}^2$ let 
$\tau(0,z)$ be the probability that the origin $0$ is 
occupied
and the cluster containing it also contains $z$. It is
further suggested that
$$
  \tau(0,z)\sim \vert z\vert^{-\eta},\qquad \eta=5/24.\tag 
2.5b
$$
This is the $\eta$ that we want to define as a crossing
probability.

 Let $d$ be large but small in proportion to $\vert z 
\vert$,
and for simplicity take $z=(x,0)$ with $x>0$. Since we
shall be applying the notions of conformal
invariance we treat $z$ as a point in the complex plane.
To estimate the probability $P(z,d)$ that $0$ is occupied
and that the cluster containing it meets the disk
of radius $d$ about $z$, we apply a conformal
transformation $\phi$ that takes this disk to the exterior
of a circle of radius $R$, and has derivative equal to $1$
at the point $0$.
(It is natural to assume that 
conformal invariance is applicable to events
involving
points only if the scale at the
points is preserved.)
Since the scale is preserved at $0$,
conformal invariance
suggests that
$$
  P(z,d)\sim P(R)\sim R^{-1/\rho}.
$$

 At this level of argument, it is not worthwhile to search
for the precise formula for $\phi$. The approximation
$$
  \phi:w\rightarrow \frac{xw}{x-w}\tag 2.5c
$$
is sufficient. It takes the origin to the origin, and the
circle of radius $d$ about $x$ to the circle with center
on the real line that contains both $-x(x+d)/d$ and
$x(x-d)/d$. Thus $x^2/d$ is a fair approximation to
$R$, and
$$
   P(x,d)\sim\left( \frac{x^2}{d}\right)^{-5/48}.
$$
Thus
$$
  P(x,d)\sim\tau(0,z)/d^{-5/48}.
$$

 Now choose two large numbers $d_1$ and $d_2$, small
in proportion to $x$, and consider the probability
$P(z,d_1,d_2)$ that there is a cluster that meets
both the disk of radius $d_1$ about $0$ and the
disk of radius $d_2$ about $z$. Symmetry suggests
that 
$$
  P(z,d_1,d_2)\sim \left( \frac{x^2}{d_1d_2}\right)^{-5/48}.
$$

 On the other hand the mapping (2.5c) takes the region 
outside
the two disks about $0$ and $z$ to the
annular region between two circles of radii
about $d_1$ and $x^2/d_2$  and with centers close to $0$.
We conclude that the probability of a crossing
from one side to another of an annulus with center
$0$ and radii
$r_1<r_2$
is approximately
$$
  \left(\frac{r_2}{r_1}\right)^{-5/48}.
$$
This relation is confirmed by numerical simulations
that we do not present and that were
much less systematic than those of Part Three.
It yields a definition of $\eta$ in terms of the
crossing probabilities for an annulus.

\fighere{11 pc}\caption{Figure 2.5. \rm The map used to 
define
the exponent $\eta$ in the finite models. (The radial 
scale of
the second drawing is logarithmic.)}

 In the numerical studies (\cite{L2}) of finite models,
no attempt has been made to determine an
approximate value for $\eta$. The procedure
that might be used is clear.
Suppose that, as in \S 2.3, we define the finite model
by a decomposition of the sides of a square into $l$
intervals of equal length. The map $\Theta$ was defined
by juxtaposing four such squares into a $2\times2$ array.
If $m$ and $n$ are two integers,
we can also juxtapose $mn$ squares to form
an $m\times n$ array.
The definition of $\Theta$ can
be extended to give crossing probabilities
between intervals of length $1/l$
in the resulting rectangle of base $n$ and height $m$.

 The function 
$$
   \exp(\tfrac{2\pi(z+1)}{m})\tag 2.5d
$$
takes the rectangle of base $\{0,n\}$ and side
$\{0,im\}$ to the annulus of
radii $\exp(2\pi/m)$ and $\exp(2\pi(n+1)/m)$. 
Provided $m>1$ the annulus is thus represented as
the glueing of $mn$
conformally distorted squares, as in Figure 2.5, 
and the definition
of $\Theta$ could 
be mimicked to define at a finite
level the probability of crossing an annulus.

\subheading{\num{2.6.} Conformally invariant fields and 
percolation}
In response to Aizenman's suggestion of conformal invariance
Cardy \cite {C4} proposed, on the basis of the theory
of conformally invariant fields, a formula for 
the horizontal crossing probability $\pi_h(r)$ on rectangles
of aspect ratio $r$. In other words, if one takes $E$ to 
be defined
by a rectangular curve $R$ of width $a$ and height $b$ 
such that 
$r=a/b$ and
by opposing horizontal sides $\alpha,\beta$ 
with no excluded crossings, then a formula for 
$\pi(E,M_0)$ can be obtained
that is confirmed by the numerical results of \cite {U} 
and of  
\S 3.2 below.
The coincidence of the predicted values with those found by
simulation is the strongest evidence yet for
conformal invariance. We stress nonetheless\ that
the conformal invariance for events $E$ other than those 
defined
by a single pair of intervals is not yet, even 
conjecturally,
a consequence of the theory of conformally invariant
fields.

 To give two intervals on the simple closed curve $C$ is 
to give four
points $z_1$, $z_2$, $z_3$, and $z_4$ in clockwise order.
The first two are the endpoints of $\alpha$ and the last two
the endpoints of $\beta$. There is 
a conformal (holomorphic) map of
the interior of $C$ to the unit disk that takes $C$ to the 
circumference
and $z_1,z_2,z_3,z_4$ to four points $w_1,w_2,w_3,w_4$.
The map is not uniquely determined, for it can be followed
by any conformal automorphism of the disk. Only the 
cross-ratio
$$
   \frac{(w_4-w_3)(w_2-w_1)}{(w_3-w_1)(w_4-w_2)}
$$
is uniquely determined. It is a real number
between $0$ and $1$. Thus we may choose, and it is
convenient to do so, the four points $w_i$ so that
$w_1=w_0=\exp(i\theta_0)$, $w_2=\bar w_0$, $w_3=-w_0$, and 
$w_4=-\bar w_0$. Then the
cross-ratio is $\sin^2(\theta_0)$. Observe that $0\leq 
\theta_0\leq\tfrac{\pi}2$
or $\pi\leq\theta_0\leq\tfrac{3\pi}2$. Interchanging 
$\alpha$ and
$\beta$ if necessary, we usually assume the first 
alternative.

If $E$ is the event defined by
the rectangle $R$, $\alpha$, and $\beta$,
then
Cardy's formula for $\pi(E,M_0)$ is
$$
  \pi(E,M_0)=
     \frac{3\Gamma(\tfrac23)}{\Gamma(\tfrac13)^2}\sin^{%
\tfrac23}(\theta_0)\,
       {}_2\!F_1(\tfrac13,\tfrac23,\tfrac43,\sin^2(%
\theta_0)).\tag 2.6a
$$
This is a function that equals $0$ when $\theta_0=0$
and $1$ when $\theta_0=\tfrac\pi2$, as it should.

 In this paragraph we review the essential ideas
of the derivation, which is not rigorous.
Although the lattice models of statistical mechanics,
their scaling limits, and conformally
invariant field theories are
objects that can be introduced in strictly mathematical
terms, they arise,
as we saw in \S 2.2, in a physical context
rich in experience and inspiration whose
sources of insight are unfamiliar to the mathematician,
and of difficult access, so that,
intimidated and sometimes at sea, he hesitates to apply 
his usual criteria.
Our presentation of the ideas
leading to Cardy's formula (2.6a) suffers from
the attendant ambivalence; the authors have not all
persuaded themselves that they fully comprehend to
what extent the arguments are 
formal, inspired by the
physical and historical connotations of the symbols,
and to what extent they involve precisely defined
mathematical entities. As stressed in the
introduction, this section is not necessary
to the understanding of Part Three.

 In planar lattice models of statistical mechanics
such as the Ising model a state $s$, before passage to the
bulk limit, is described by its values at the sites
of the lattice that lie in some large square.
The interaction between the various points determines
the energy $H(s)$ of the state, and its Boltzmann weight
$\exp(-\beta H(s))$. The constant $\beta$, in essence the
inverse temperature, may for our purposes be taken
equal to $1$. The very important partition function is
$$
    Z(\beta)=\sum_s\exp(-\beta H(s)).
$$
It is used in particular to normalize the Boltzmann 
weights and thereby define
a measure on the set of states,
$$
     \mu(s)=\frac{\exp(-\beta H(s))}{Z(\beta)}.
$$
The natural functions of which to take expectations
$E(f)$ are those that depend on the values $s(P)$
of the state at a finite number of points. For such
a function one can expect that $E(f)$ continues to
exist in the bulk limit.

 The passage from the probabilistic concepts
of statistical mechanics to a field theory can be 
presented rigorously as an analogue of that from a 
one-parameter semigroup to the associated infinitesimal 
generator
(\cite{GJ}); in practice, however, it is
a much more adaptable and unconstrained mechanism.

 For percolation, the procedure, quite apart from questions
of the existence or nature of limits,
does not appear promising. A state $s$ is determined by the
occupied sites; the others are unoccupied. If their
number is $N(s)$ then
$$
H(s)=\{-\ln p+\ln(1-p)\}N(s)
$$ 
and
the Boltzmann weight is
$$
    \exp(-H(s))=\big(\frac{p}{1-p}\big)^{N(s)}.
$$
The value of the partition function is $(1-p)^{-N}$
if $N$ is the total number of sites in the square,
and the probability of $s$ is $p^{N(s)}(1-p)^{N-N(s)}$.

 These are the probabilities familiar from percolation, in 
which the value
of the states at the sites are independent of each other.
Thus if $f_P$ is a function of states given by
$$
      f_P(s)=f(s(P)),
$$
the function $f$ being a function on the set of possible 
values, then
for $r$ sites different from each other
$$
   E(f_{P_1}f_{P_2}\dots f_{P_r})
     =E(f_{P_1})E(f_{P_2})\dots E(f_{P_r})=E(f)^r.
$$
Passing formally to operators and to limits, we see that
$$
   E(f_{P_1}f_{P_2}\dots f_{P_r})
    =\langle\,\vert 
\phi(P_1)\phi(P_2)\dots\phi(P_r)\vert\,\rangle,
$$
if $\phi(P)=\phi$ is constant and simply 
equal to a scalar $E(f)$
operating on a space of dimension one.
Such trivial operators will not help in finding
a formula for $\eta_h$, but these considerations
do suggest that the central charge $c$ is $0$
for percolation. 

  The statistical mechanics of lattices in
a half space, or any bounded region, has, however, 
features that differentiate it from
the theory in the full space.
Boundary conditions have a much stronger effect; so familiar
uniqueness theorems for Gibbs states and correlation
functions need no longer apply.
The consequences may continue to manifest
themselves in the scaling limit.
Cardy had pointed out in \cite{C1} that
at criticality and in two dimensions 
the limit could continue to 
exhibit conformal invariance,
although of a somewhat different nature 
than for the scaling limit of bulk theories. 
In \cite {C2} and \cite{C3} he examined
the effect of modification of the boundary conditions
at the surface on the correlation
functions in the interior.

 From the principles \cite{BPZ} that 
prescribe the behavior of
conformally invariant fields in the full plane, we cite 
two. 
The first,
a global principle, is that, if 
$P$ is treated as a complex parameter
$z$, the correlation functions
$$
   \langle \,\vert\phi(P_1)\dots\phi(P_r)\vert\,\rangle
$$
may be treated as analytic functions of $z_1,\dots,z_r$ 
and of
their complex conjugates $\bar z_1,\dots,\bar z_r$ and as 
such
transform in a prescribed
way under holomorphic
(and antiholomorphic) maps $w(z)$. The simplest relation
appears for the fields called primary:
$$
    \langle\,\vert\phi(z_1,\bar z_1)\dots\phi(z_r,\bar 
z_r)\vert\,\rangle=
    \prod w'(z_i)^{h_i} \bar w'(\bar z_i)^{\bar h_i}
    \langle\,\vert \phi(w_1,\bar w_1)\dots\phi(w_r,\bar 
w_r)\vert\,\rangle,
$$
where the $h_i$ and $\bar h_i$ are known 
as the conformal dimensions of 
the field $\phi_i(z_i,\bar z_i)$.

 At each point $P=z$, 
we may consider the algebras of formal
holomorphic and antiholomorphic
vector fields defined in a complement of the point
(more precisely central extensions,
the Virasora algebras, of these
two algebras). The second principle is that there
is an action of these algebras on the spaces underlying the
fields and
on the fields of operators themselves.
There are conditions of compatibility,
but they are subtle.

 Conformally invariant fields
are introduced in order to describe the asymptotic
behavior at large differences of correlation
functions of field theories,
either on a lattice or in the continuum, in the sense
of \cite{GJ}, so that it is perhaps ingenuous
to expect them to have the same kind of operator
significance.
They are defined by Laurent series in which the
individual coefficients are meaningful objects; thus
they can be integrated against a limited class of functions
on appropriate curves surrounding the point under 
consideration.
Since the theory is conformally
invariant, one could pass to the Riemann sphere 
and take this curve to be
the image of a straight 
line in the plane, thereby recovering more familiar
objects, but this seems to us to do
violence to the spirit of the subject.

 In 
two dimensions a simple choice of
half space
is the upper half-plane, with the real axis as boundary,
and in this
context there are further
principles
\cite {C3, pp. 584-585} that are not at all obvious,
at least to us; indeed we are not at all confident
that we have adequately comprehended Cardy's views.
The principles need nevertheless to be stressed. 

 A first, patent, principle is that the relevant
algebra is not the sum of the holomorphic and
antiholomorphic algebras, but the diagonal algebra
contained therein, for the real axis, as the boundary of the
region, must be left invariant.

Secondly, there are two pertinent classes of boundary
conditions with
quite different properties, those that are translation 
invariant,
thus homogeneous on the entire boundary,
and those that are homogeneous on both
sides of $0$
(so that scaling is still meaningful) but differ from one 
side 
to the other.

For those that are homogeneous on the entire line, it 
appears
not unreasonable to expect that the underlying spaces are
direct sums of irreducible representations of the Virasoro 
algebra,
although the possibility of imposing different homogeneous
boundary conditions may entail a rich variety of {\it 
sectors}
in these sums. We do not yet understand
to what extent other representations
than the trivial one are necessary
for percolation
with homogeneous
boundary conditions (whatever these might be!).
For a boundary condition with a transition at $0$
the representations of the Virasoro algebra need
not be irreducible. 
The vacuum associated to these boundary conditions
is not translation invariant, and thus is not annihilated by
$L_{-1}$.

It appears that the sector
(or theory, or, more concretely, the
underlying Hilbert space\,---\, it is a matter
of terminology) defined 
by such boundary conditions
can be obtained from the full
homogeneous sector by applying an operator $\phi=\phi(0)$.
Once we have identified the boundary operators, and
persuaded ourselves of the
conformal invariance, so that
the operators depend on a parameter $z$, they can be used
to insert boundary conditions at several points.

 We have already remarked that the first representation
of the Virasoro algebra that
appears in the study of percolation is the trivial
representation. Useless though it appears to be for the
study of correlation functions, it did yield immediately
the value $0$ for the central charge $c$.

 The primary boundary operator $\phi(0)$ acting on the 
vacuum $\vert\,\rangle$
will yield the vacuum $\phi(0)\vert\,\rangle$ associated
to the boundary conditions, and $\phi(0)\vert\,\rangle$
will be the highest-weight vector of a representation of the
Virasoro algebra. 
According  to 
the results of \cite {RW1, RW2} 
and authors there cited, an example of a
representation that has the trivial representation as
a quotient but for which the highest-weight
vector is not translation invariant is
obtained by dividing the Verma module with parameters
$c=0$ and $h=0$ by the submodule with parameters
$c=0$ and $h=2$. Since this is the submodule generated
by the null vector corresponding to the root
$$
    h_{1,2}=\frac{((m+1)-2m)^2-1}{4m(m+1)}=0,\qquad m=2,
$$
of the Kac determinant formula, Cardy writes
$\phi_{1,2}$ rather than  
$\phi$. 

 This argument, however, is far from satisfactory, for we 
have
not even been precise about the nature of the 
boundary conditions. Cardy's argument 
draws on more sources. In particular,
it exploits a common, but entirely factitious,
device for introducing boundary conditions into
percolation by treating it as a degenerate
case of the $q$-state Potts model. The device
has the additional advantage that the crossing probabilities
appear as correlation functions.

 Recall \cite {W} that the Potts model is a lattice model, 
in which there are $q\geq1$
possible values $\sigma$
for a state at each site of a square lattice. The 
hamiltonian
is
$$
 H(\sigma)=\sum_{x,y} 1-\delta_{\sigma_x,\sigma_y}.
$$
The sum runs over all pairs of nearest neighbors inside a 
large square
laid over the lattice. Observe that the extra term $1$ 
does not affect
the Boltzmann weights.
In contrast to percolation, when $q>1$ there is a genuine 
energy of
interaction. 

 Let ${\frak B}$ be the set
of nearest-neighbor bonds. The partition function for free 
boundary conditions
is obtained by summing
$$
  \exp(-\beta H(\sigma))
 =\prod_{\{x,y\}\in \frak B} (e^{-\beta}+
(1-e^{-\beta})\delta_{\sigma_x,\sigma_y}).
$$
Setting $p=1-e^{-\beta}$ we may write this as
$$
     (1-p)^d,
$$
with $d$ equal to the number of bonds joining two sites 
with $\sigma_x
\neq\sigma_y$. We may also write it as
a sum over the subsets  $\frak x$ of $\frak B$,
$$
   \sum p^{B(\frak x)}
  (1-p)^{B-B(\frak x)}\prod_{\{x,y\}\in\frak 
x}\delta_{\sigma_x,\sigma_y}.
   \tag 2.6b
$$
\noindent
\looseness=-1
The \kern-.42pt integer \kern-.42pt $B$ \kern-.42pt is 
\kern-.42pt the
\kern-.42pt total \kern-.42pt number \kern-.42pt of 
\kern-.42pt bonds
\kern-.42pt and \kern-.42pt $B(\frak x)$
\kern-.42pt the \kern-.42pt number \kern-.42pt of 
\kern-.42pt bonds \kern-.42pt
in \kern-.42pt $\frak x$.

 Each subset  $\frak x$ of $\frak B$ decomposes the set 
$\frak S$ of sites
into clusters, two points lying in the same cluster if
they can be joined by a sequence of bonds in $\frak x$.
The product 
$$
\prod_{\{x,y\}\in\frak x}\delta_{\sigma_x,\sigma_y}
$$
is $0$ or $1$, and is $1$ if and
only if $\sigma$ is constant on each cluster. We write
$\frak x\rightarrow \frak A$ if $\frak A$ is the 
family of clusters
determined by $\frak x$. The clusters in $\frak A$ are
denoted $A_1,\dots,A_r$. The integer $r$ is equal to the
number $N(\frak A)$ of clusters in $\frak A$. The sum
(2.6b) is also equal to a sum over all possible 
decompositions
into clusters, 
$$
  \sum_{\frak A}\sum_{\frak x\rightarrow \frak A}
   p^{B(\frak x)}
   (1-p)^{B-B(\frak x)}\prod_i\prod_{\{x,y\}\in 
A_i}\delta_{\sigma_x,\sigma_y}.
$$
Taking the sum over all states, we find, as in \cite{E1, 
\S 2.2}, 
that the partition function
with free boundary conditions is equal to
$$
  Z_f= \sum_{\frak A}\sum_{\frak x\rightarrow \frak A}
   p^{B(\frak x)}
   (1-p)^{B-B(\frak x)}q^{N(\frak A)}.\tag 2.6c
$$

 To examine the effect of boundary conditions we consider a
rectangle, imposing boundary conditions on the left and 
right sides
but leaving the top and bottom free. Suppose we demand that
$\sigma$ take only the value $\alpha$ on the left side
and only the value $\beta$ on the right side. Then the 
partition function
is
$Z_{\alpha,\beta}$  and it is obtained from (2.6c) 
on replacing $N(\frak A)$ by the number 
$N'(\frak A)$ of clusters that do not
intersect the left or right sides. Moreover, if 
$\alpha\neq\beta$
then all families
of clusters with a member that meets the left and right
sides are excluded from the sum. Consequently
the difference
$$
   Z_{\alpha,\alpha}-Z_{\alpha,\beta}\qquad\alpha\neq%
\beta,\tag 2.6d
$$
is equal to the sum of the expression
$$
  \sum_{\frak x\rightarrow \frak A}
   p^{B(\frak x)}
   (1-p)^{B-B(\frak x)}q^{N'(\frak A)}.
$$
over those families of clusters that do contain
a member that intersects both sides of the square.

In particular, setting formally $q=1$ we obtain
the sum over all subsets $\frak x$ of the set of bonds
that admit a horizontal crossing of
$$
 p^{B(\frak x)}(1-p)^{B-B(\frak x)}. 
$$
When $p$ is the critical probability for bond percolation
this is the probability of a horizontal crossing, thus in
essence $\pi_h$.

 We have progressed in two ways. First of all, the 
crossing probability
$\pi_h$ has been identified as a 
difference of partition functions, and thus, as we shall 
see,
as a difference of correlations. Secondly there is a
free parameter $q$ and with a little bit of courage,
we can transfer results for $q>1$ to $q=1$. That the
condition $\alpha\neq\beta$ can not be realized for
$q=1$ will trouble only the fainthearted, for it will never
explicitly
enter our manipulations of (2.6d).

 What is relevant in (2.6d) is that 
the expression is a linear combination
of partition functions with boundary 
conditions that change
at four points, the four corners of the rectangle, 
from fixed to free. Although the transition from
partition functions to correlation functions
appears to be more a matter of intuition than
of logic, persuasive only after much experience with 
the passage from lattice models to operators, it 
does appear rather explicitly in Cardy's reflections
\cite{C3, pp. 584-585}
for the case of a transition from a homogeneous condition
$\sigma(x)=\alpha$ to the condition
$\sigma(x)=\alpha$ for  $x<0$ and
$\sigma(x)=\beta$ for $x>0$. The corresponding operator
is denoted somewhat informally as $\phi_{\alpha,\beta}$
or $\phi_{\alpha,\beta}(0)$. We suppress from the notation
that there is also a jump in the boundary conditions at 
$\infty$, and 
of course admit the possibility that $\alpha$ signifies
a free boundary condition, as well as a definite value of
the spin or other variable. 

 In the context of conformally invariant theories it is 
possible
to use the transformation
$w=\ln z$ to replace the upper half-plane,
with the point $0$ on the boundary distinguished, by the
strip $0\leq\Im w\leq\pi$. 
Translation-invariant boundary conditions
are transferred to 
boundary conditions equal on both sides,
and translation invariant with
respect to the strip. Boundary conditions
with a jump are transferred to boundary conditions
different on both sides of the strip, but
translation invariant with respect to it.
Experience with 
limits of standard
lattice models, above all the Ising model, makes clear
that calculating partition functions
and correlation functions, or rather their limits,
on such strips with boundary conditions at $\Im w=0$
and $\Im w=\pi$ is above all a matter of calculating the
eigenvector $v_{\alpha,\beta}$ associated 
to the smallest eigenvalue of
the transfer matrix
associated to these conditions.
If $\vert\,\rangle$ is the eigenvector associated to equal
homogeneous boundary 
conditions and $\phi_{\alpha,\beta}$ is an operator taking 
$\vert\,\rangle$ to the eigenvector $v_{\alpha,\beta}$
then a
correlation function 
$$
   \langle\,\vert\phi_1\dots\phi_r\vert\,\rangle
$$
is replaced by
$$
   \langle\,\vert\phi^*_{\alpha,\beta}
     \phi_1\dots\phi_r\phi_{\alpha,\beta}\vert\,\rangle
    =\langle\,\vert\phi_{\beta,\alpha}     
\phi_1\dots\phi_r\phi_{\alpha,\beta}\vert\,\rangle.\tag 
2.6e
$$
However we have implicitly allowed a jump in the boundary 
conditions
at $0$ and at $\infty$, so that, indicating the dependence
of one of the operators on the point $0$ and the other
on the point $\infty$,
this equation
might be rewritten as 
$$
    \langle\,\vert\phi_{\beta,\alpha}(\infty)
     \phi_1\dots\phi_r\phi_{\alpha,\beta}(0)\vert\,%
\rangle.\tag 2.6f
$$

 Transforming
back to the upper half plane, and allowing insertions
of modifications at several, say four, points, one of 
which may be 
at infinity, we obtain, for $r=0$,
$$
  \langle\,\vert\phi_{\alpha,\beta}(z_1)\phi_{\beta,%
\gamma}(z_2)
     \phi_{\gamma,\delta}(z_3)\phi_{\delta,\alpha}(z_4)%
\vert\,\rangle.\tag 2.6g
$$
If $r>0$ it is less clear where to insert the operators
in (2.6e), but $r=0$
is the pertinent value of $r$ for this is the value for
which (2.6e) is a partition function.
Although the modification in the boundary values was 
taken to be from
one prescribed value to another and not
from a prescribed value to free
boundary conditions, the same arguments are 
valid in both cases. It is 
the transition from free to fixed, 
$\phi_{f,\alpha}$, and from fixed to free,
$\phi_{\alpha,f}$, that appear in (2.6d) because the
one pair of sides on which the boundary conditions are
fixed are separated by the other sides on which they are 
free. 

 Cardy \cite{C1, C2, C3, C4} does not find the operators 
$\phi_{\alpha,f}$
directly. Rather he argues first (for $q>1$ but also by
extrapolation for $q=1$) that the operator
$\phi_{\alpha,\beta}$ associated to the transition from 
one fixed boundary condition to
another, different, fixed condition is 
the primary field $\phi_{1,3}$, and then
that the operator-product expansion
of $\phi_{\alpha,f}(z)\phi_{f,\beta}(w)$,
which would be
$$
   \phi_{\alpha,f}(z)\phi_{f,\beta}(w)
    \sim \delta_{\alpha,\beta}\text{\bf 1}+
\phi_{\alpha,\beta}
$$
implies that
$$
   \phi_{\alpha,f}=\phi_{1,2}.
$$
Since his final argument is somewhat more convincing for
unitary theories than for non-unitary theories, it
is again best to regard it as extending to $q=1$
by extrapolation.

 The identification of the operators $\phi_{\alpha,\beta}$
appeals to experience with specific models that,
like the operator-product expansion itself, may be
unfamiliar to the mathematician; so we
observe that
the numbers $\pi(E,M)$ are, by their very
definition, invariant under
dilations of the data defining $E$. In particular, if (2.6g)
is to represent a probability of crossing between intervals
defined by $z_1$, $z_2$, $z_3$, and $z_4$ then it
must be homogeneous of degree $0$ in
the vector $(z_1,z_2,z_3,z_4)$. Since the operator
$\phi_{\alpha,f}(z)=\phi_{f,\alpha}(z)$ is primary it is 
homogeneous of some degree $h$, and $h$ must
be $0$.

 Although, in principle, any 
positive real number $h$ is a possible degree of 
homogeneity,
those that occur most commonly are those associated 
to reducible Verma modules, and these are given
by the 
Kac formula, which at $c=0$ becomes
$$
   h_{p,q}=\frac1{24}((3p-2q)^2-1),
$$
where $p$ and $q$ are positive integers. The simplest
choices of $p$ and $q$ that give $h=0$ are $p=q=1$,
which leads to the trivial representation,
and $p=1$, $q=2$, that yield $\phi_{\alpha,f}=\phi_{1,2}$.

 To complete the derivation of Cardy's formula, we use
the ideas of \cite{BPZ} as presented in \cite{SA} to find
the differential equation satisfied by (2.6g).
The null vector  $v_{1,2}$ in the Verma module with 
parameter
$c=1-6/m(m+1)$ is
$$
     (L^2_{-1}-\frac13(4h_{1,2}+2)L_{-2})\vert 
h_{1,2}\rangle.\tag 2.6h
$$
where $\vert h_{1,2}\rangle$ is the highest weight 
vector of the Verma
module.
For $c=0$ and $m=2$, $h_{1,2}=0$.
Moreover, according to formula (4.6.21)
of \cite {SA}, to find the differential equations
satisfied by (2.6g) we replace
$L_{-k}$ in (2.6h) by
$$
   {\Cal 
L}_{-k}=-\sum_{i=1}^{3}\frac1{(z_i-z_4)^{k-1}}\partial_i,
$$
an expression that the relation $h_{1,2}=0$ has made 
much simpler than it would otherwise be.

 The translation invariance permits the replacement of
$$
  -\sum_{i=1}^3\partial_i
$$
by $\partial_4$, so that the differential equation satisfied
by (2.6g) is
$$
  \big(\partial_4^2+\frac23(\frac1{z_3-z_4}\partial_3+
\frac1{z_2-z_4}\partial_2+\frac1{z_1-z_4}\partial_1)
    \big)\langle\dots\rangle=0.\tag 2.6i
$$
If we set
$$
   z=\frac{(z_1-z_2)(z_3-z_4)}{(z_1-z_3)(z_2-z_4)},
$$
then conformal invariance implies that (2.6g) is a function
$g$ of $z$ alone.

 With a little effort 
we infer from (2.6i) that $g$
satisfies the equation
$$
   z(1-z)^2g''+2z(z-1)g'+\frac23g'-\frac23z^2g'=0,
$$
or upon simplification
$$
  z(1-z)g''+\frac23(1-2z)g'=0.
$$
This is a degenerate hypergeometric equation with
two solutions $g\equiv1$
and
$$
  g(z)= 
z^{\tfrac13}{}_2\!F_1(\frac13,\frac23,\frac43;z).\tag 2.6j
$$

 To determine which 
linear combination of these two 
solutions is pertinent to our problem, we take
$z_1$, $z_2$, $z_3$, and $z_4$ in decreasing order to be 
the images
of the four vertices of the rectangle in clockwise order, 
starting
with the lower left corner.
If $r$ is the aspect ratio of the rectangle 
then $z\rightarrow 0$ when $r\rightarrow\infty$ and
$z\rightarrow 1$ when $r\rightarrow 0$. Thus the solution 
yielding 
the crossing probability $\pi_h(R_r, M_0)$ must be a 
constant times
(2.6j). The identity
$$
   \frac{3\Gamma(\tfrac23)}{\Gamma(\tfrac13)^2}
   z^{\tfrac13}{}_2\!F_1(\frac13,\frac23,\frac43;z)=
     1-\frac{3\Gamma(\tfrac23)}{\Gamma(\tfrac13)^2}
   (1-z)^{\tfrac13}{}_2\!F_1(\frac13,\frac23,\frac43;1-z)
$$
implies that the constant must be
$$
   \frac{3\Gamma(\tfrac23)}{\Gamma(\tfrac13)^2}
$$
in order that the function have the
correct behavior at $z=1$. This is the formula (2.6a) of 
Cardy
in a different notation (and for the upper half-plane rather
than the unit disk.)

\heading 3. The experiments\endheading

\subheading{\num{3.1.} Experimental procedure}
In order to provide some evidence for the hypotheses
of universality and conformal
invariance, we performed several
simulations. Although several artifices had to be used in 
the various cases,
the basic method is the same throughout: ({\sl i}) draw 
the curve $C$ defining
the event $E$ on the lattice, ({\sl ii}) assign randomly 
to each site of the
lattice lying inside the curve a state (open with 
probability $p_c$, closed
with 
probability $(1-p_c)$) and ({\sl iii}) check whether the 
various crossings defining
the event $E$ exist or not. These three steps are repeated 
till the desired 
sample size is reached. The estimated value of $\pi(E)$,
denoted $\hat\pi(E)$, is then the ratio of the number
of configurations satisfying the conditions of $E$ to the 
sample size. 

For the above experimental procedure, the statistical 
errors are the easiest
to assess. The sample size for all our experiments was at 
least $10^5$, and
very often larger. For an estimated value $\hat\pi\sim 
0.5$, this leads
to a statistical error of $\Delta\hat\pi\sim 3\times 
10^{-3}$.  For
the largest $\hat\pi$ measurable ($\sim 0.999$)
or\ the smallest ($\sim 0.001$),
the error is $\sim 2\times 10^{-4}$. (All
statistical errors are taken to represent a $95\%$ 
confidence interval.)

The systematic errors are of various origins. Probably the 
least important
source is the random number generator. We used 
in most of the experiments
the linear congruential generator
$x_{i+1}=(ax_i+c)\mod m$, with 
$$a=142412240584757,\qquad c=11,\qquad m=2^{48}.$$
It is
of maximal period $m$. We believe it to be satisfactory.

A second source is the ``value'' of
the probability $p_c$ appearing in the statement
of the theorem of \S 2.1.
This critical value $p_c$ is a well-defined concept only for
percolation phenomena on an {\it infinite} lattice. But 
all our simulations
are carried on finite lattices! The solution to this 
difficulty calls for a
compromise. Indeed, on the one hand, lattices have to be 
chosen large enough
to give a good approximation of the infinite case. On the 
other hand, a larger
lattice requires a better approximation for $p_c$. (Recall 
that the slope of
$\pi_h$ around $p_c$ increases with the size of the 
lattice, as depicted
in Figure 2.1c.)
The most suitable 
approximation depends, as discussed
in \cite{Z} and \cite{U},
on the size of the lattice;
one could even imagine that it is different
for rectangles containing the same number of sites but
with distinct aspect ratios $r$. All the experiments but 
one were conducted
at $p_c=0.59273$ with the curves $C$ containing from 
40,000 to 200,000 sites.
The only experiment that used a different $p_c$ was a 
repetition of the
principal experiment of \cite{U} where we measured the 
universal functions
$\eta_h$, $\eta_v$ and $\eta_{hv}$ to be defined below.
As these data together
with Cardy's prediction are to be used as yardsticks for 
the new experiments,
we felt that measuring them on a larger lattice was 
appropriate. For that
experiment on a larger lattice,
we used $p_c=0.5927439$. The six first digits in $p_c$ were 
definitely necessary to achieve the desired precision. 
The results are discussed
in paragraph 3.2.   

Another important source of systematic errors is the 
convention of crossings
on finite lattices. The curve $C$ is to be drawn
in the plane containing
the lattice. We chose to define a crossing from the 
interval $\alpha$ to
$\beta$ on $C$ as starting from a site inside $C$ joined 
to a neighbor by
a bond intersecting the image of $\alpha$ of the lattice. 
Similarly the
crossing must end at a site inside $C$ such that one of 
the attached edges
intersect the image of $\beta$. Note that we might well 
have defined the
crossing as starting from an open site {\it outside} the 
curve $C$ with one
attached edge intersecting the interval in question. Hence 
the convention used
introduces a systematic error. Moreover, one can imagine 
easily that sliding
rigidly a rectangle on a square lattice by a fraction of 
the mesh might
add a whole line or column to the set of inner sites,
thus changing the estimate
$\hat\pi$. 
For reasons described in
\cite{U}, the attendant error 
for rectangles is commensurate with ${2\over n}\pi'$ 
where $\pi'$ stands for the derivative of $\pi$ with 
respect to 
the aspect ratio $r$ and $n$ is the linear dimension of 
the rectangle.
For a square containing $200\times200$ sites, the error on 
$\hat\pi_h$ turns
out to be $\sim 5\times 10^{-3}$, larger than the 
statistical error introduced
by a sample of 100,000 configurations. We were
on the whole content if the results obtained by simulation
were consistent with those predicted by
universality and conformal invariance within five parts in 
one thousand.

 For the final experiments on conformal invariance,
it grew slightly larger than one part in
one hundred.
This is not surprising in view of
further specific sources of systematic errors,
due to penetrating angles, branch points, 
and unbounded regions, that will be discussed as they arise.
It does nevertheless make
further experiments desirable.

The events studied in \cite{U}
were defined by a rectangular curve $C$.
We chose the collection of intervals in four different ways.
First of all,
$\alpha$ could be the left side of the rectangle and 
$\beta$ the right,
which yielded the probability $\pi_h(M)$ of a horizontal 
crossing,
or $\alpha$ the lower side and $\beta$ the upper, which 
yielded  
the probability $\pi_v(M)$ of a vertical crossing. We also 
studied
the probability $\pi_{hv}(M)$ of simultaneous horizontal and
vertical crossings. The difference $\pi_{h}(M)-\pi_{hv}(M)$
provides an example of an event with a horizontal crossing 
but no vertical
one. In the notation of \S2.3, the intervals 
$\alpha, \delta, \beta, \gamma$ 
are then the left, upper, right and lower sides, 
respectively.
For a little variety the probability
$\pi_d(M)$ of a crossing from the upper half of the left 
side to the 
right half of the bottom side was also studied.
In these functions there is an implicit variable $r$, the 
aspect ratio
of the rectangle, that we took to be the quotient
of the length of the horizontal side by 
that of the vertical side.
Taking $M$ to be $M_0$ we obtain, as explained
in \cite{U}, 
four universal functions, $\eta_h=\pi_h(M_0)$, 
$\eta_v=\pi_v(M_0)$,
$\eta_{hv}=\pi_{hv}(M_0)$ and $\eta_d=\pi_d(M_0)$ of $r$.
The
probabilities of similar events will be measured in some
of the following experiments.

If the hypothesis of universality holds, the functions 
$\eta_h(r)$ and 
$\eta_v(r)$ are not independent. This can be seen by the 
following 
duality argument. We draw a rectangle 
on a {\sl triangular} lattice. There will be a horizontal 
crossing (on open
sites) if and only if there is no vertical crossing on 
{\sl closed} 
sites. This is consistent with the theorem of \S 2.1 
only if $p_c={1\over2}$
for this model  (denoted $M$) and then
$$
\pi_h(r,M)+\pi_v(r,M)=1
$$
for all $r$. Of course, the argument could have been made 
for any closed
curve $C$, disjoint intervals $\alpha$ and $\beta$, and 
the two disjoint
intervals $\delta$ and $\gamma$ of their complement. The 
relation would then
be 
$$
\pi_{\alpha\leftrightarrow\beta}+
\pi_{\delta\leftrightarrow\gamma}=1
$$ 
where
$\pi_{\alpha\leftrightarrow\beta}$ stands for the 
probability of a crossing
from $\alpha$ to $\beta$.  Universality then 
implies that this relation holds for any
model.  
This is a handy test of simulations. Observe that, for the 
model
$M_0$, the complementarity of horizontal crossings on open 
sites and of
vertical crossings on closed sites {\sl does not} hold
for individual configurations. Every experiment 
measuring simultaneously $\pi_h(M)$ and $\pi_v(M)$ on 
other models
$M$, such as 
$M_0$, for which this complementarity does not hold 
for direct reasons serves as a check on 
universality.

In the tables, the results of the experiments are presented 
together with
either Cardy's prediction when it is applicable or by values
for rectangles
inferred by interpolation
from the experimental results
of the next section.
Cardy's prediction will be denoted by $\pi^{cft}$
for {\it conformal field theory} and the estimated values 
for rectangles,
as well as values calculated from them using 
interpolation, by
$\hat\pi^\square$.

\subheading{\num{3.2.} Experimental verification of 
Cardy's formula}
The goal of the first experiment is twofold: to 
verify
again Cardy's prediction for
the function $\pi_h(r)$ on $M_0$ and
to obtain values of $\pi_{hv}(r,M_0)$ suitable for
comparison in other experiments. A similar experiment was 
performed and
reported in \cite{U}
before Cardy proposed his formula. Here we increase the 
number of sites
inside the rectangles from
the approximately
40,000 used in \cite{U} to 1,000,000
and the sample size to $10^6$ configurations.
For the reasons explained above, $p_c$ was
taken to be 0.5927439. (This value
compares well with the conclusions
of Ziff \cite{Z} that came to our attention after 
the
first version of the paper was prepared.) The 
results, tabulated in Table 3.2, are a replacement
for those of Table III of \cite{U}  
and are suitable for calculating the values 
$\hat\pi_{hv}^\square$
by interpolation.

If one uses $s=\ln r$ instead of $r$, the function 
$\ln(\pi_h/(1-\pi_h))$ is odd
because of the relation $\pi_h+\pi_v=1$. The estimated 
values of this function
(dots) are plotted against Cardy's prediction (continuous 
line). The values
$\hat\pi_v$ were used for $r<1$. For $s\sim0.$, the 
measured values of
$\ln(\pi_h/(1-\pi_h)$ carry a statistical error
of $\sim4\times10^{-3}$ and for $\vert s\vert\sim2.$ this
error increases to $2\times10^{-2}$. These errors are too
small to
be indicated in Figure 3.2, being in the worst
situation approximately the size of the dots themselves.

\fighere{20 pc}\caption{Figure 3.2. \rm Comparison of 81 
measured
values of $\ln\pi_h/(1-\pi_h)$ (dots) with Cardy's 
prediction (curve).}

 A glance at Table 3.2 shows that the difference 
$\hat\pi_h-\pi^{cft}_h$ is positive
for all $r$. Though it is always $\le 6\times 10^{-4}$ and 
smaller than the
statistical error, a systematic error is suggested. 
Ziff (\cite{Z}) gave an heuristic description of the 
``good'' value of $p_c$
for the square using finite-size scaling arguments and 
simulations. 
For other curves $C$, 
including rectangles of large or small aspect
ratio, the ``best'' values
of $p_c$ for finite lattices are not available.
It may be that
the values of $p_c$ are different for the measurements of 
$\pi_h$ and of $\pi_v$
on the same rectangle, and that this is
the source of the error. Observe finally that, with this 
size of
lattice, a change of a few units in the sixth digit of $p_c$
could account for the discrepancies between $\pi^{cft}_h$ 
and $\hat\pi_h$.
Despite the systematic error, the agreement is remarkable 
and we shall compare
the results of the following experiments with $\pi^{cft}$ 
instead of 
$\hat\pi^\square$ when the former is applicable.

\pageinsert
\toptablecaption{\smc Table 3.2. \rm 
$\hat\pi_h,\hat\pi_v,\hat\pi_{hv}$
on $M_0$ for various values of the aspect ratio $r$.}
\vbox{\eightpoint
\offinterlineskip
\hrule
\halign{\strut\vrule\quad\hfil#\hfil
    \quad&\vrule\quad\hfil#\hfil\quad&
    \vrule\quad\hfil#\hfil\quad&\vrule\quad\hfil#\hfil\quad&
    \vrule\quad\hfil#\hfil\quad&\vrule\quad\hfil#\hfil\quad&
    \vrule\quad\hfil#\hfil\quad&\vrule\quad\hfil#\hfil\quad
    \vrule\cr
\hfil&\hfil&\hfil&\hfil&\hfil&\hfil&\hfil&\hfil\cr
\hfil width\hfil&\hfil height\hfil&\hfil $r$\hfil&\hfil 
$r^{-1}$\hfil&$\pi_h^{cft}$\hfil&
\hfil $\hat\pi_h$\hfil&\hfil 
$\hat\pi_v$\hfil&\hfil$\hat\pi_{hv}$\hfil\cr
\hfil&\hfil&\hfil&\hfil&\hfil&\hfil&\hfil&\hfil\cr
\noalign{\hrule height1pt}
\hfil&\hfil&\hfil&\hfil&\hfil&\hfil&\hfil&\hfil\cr   
      1000 &1000 & 1.000 & 1.0000 & 0.5000 & 0.5001 & 
0.4999 & 0.3223  \cr 
      1025 & 975 & 1.051 & 0.9512 & 0.4740 & 0.4743 & 
0.5257 & 0.3211  \cr 
      1050 & 950 & 1.105 & 0.9048 & 0.4480 & 0.4484 & 
0.5516 & 0.3180  \cr 
      1080 & 930 & 1.161 & 0.8611 & 0.4226 & 0.4230 & 
0.5768 & 0.3127  \cr 
      1105 & 905 & 1.221 & 0.8190 & 0.3970 & 0.3974 & 
0.6026 & 0.3055  \cr 
      1135 & 880 & 1.290 & 0.7753 & 0.3695 & 0.3696 & 
0.6301 & 0.2950  \cr 
      1160 & 860 & 1.349 & 0.7414 & 0.3473 & 0.3475 & 
0.6522 & 0.2854  \cr 
      1190 & 840 & 1.417 & 0.7059 & 0.3235 & 0.3236 & 
0.6762 & 0.2733  \cr 
      1220 & 820 & 1.488 & 0.6721 & 0.3003 & 0.3004 & 
0.6994 & 0.2600  \cr 
      1250 & 800 & 1.562 & 0.6400 & 0.2777 & 0.2779 & 
0.7217 & 0.2458  \cr 
      1285 & 780 & 1.647 & 0.6070 & 0.2541 & 0.2543 & 
0.7453 & 0.2297  \cr 
      1315 & 760 & 1.730 & 0.5779 & 0.2330 & 0.2333 & 
0.7666 & 0.2144  \cr 
      1350 & 740 & 1.824 & 0.5481 & 0.2111 & 0.2117 & 
0.7883 & 0.1976  \cr 
      1385 & 725 & 1.910 & 0.5235 & 0.1929 & 0.1935 & 
0.8065 & 0.1826  \cr 
      1420 & 705 & 2.014 & 0.4965 & 0.1731 & 0.1736 & 
0.8265 & 0.1657  \cr 
      1455 & 685 & 2.124 & 0.4708 & 0.1542 & 0.1546 & 
0.8450 & 0.1490  \cr 
      1490 & 670 & 2.224 & 0.4497 & 0.1389 & 0.1392 & 
0.8606 & 0.1351  \cr 
      1530 & 655 & 2.336 & 0.4281 & 0.1236 & 0.1239 & 
0.8761 & 0.1210  \cr 
      1570 & 640 & 2.453 & 0.4076 & 0.1093 & 0.1096 & 
0.8905 & 0.1077  \cr 
      1610 & 620 & 2.597 & 0.3851 & 0.09402 & 0.09424 & 
0.9055 & 0.09299  \cr 
      1650 & 605 & 2.727 & 0.3667 & 0.08201 & 0.08212 & 
0.9176 & 0.08132  \cr 
      1690 & 590 & 2.864 & 0.3491 & 0.07104 & 0.07120 & 
0.9286 & 0.07065  \cr 
      1735 & 575 & 3.017 & 0.3314 & 0.06053 & 0.06082 & 
0.9390 & 0.06047  \cr 
      1775 & 565 & 3.142 & 0.3183 & 0.05314 & 0.05332 & 
0.9463 & 0.05309  \cr 
      1820 & 550 & 3.309 & 0.3022 & 0.04459 & 0.04478 & 
0.9549 & 0.04465  \cr 
      1870 & 535 & 3.495 & 0.2861 & 0.03669 & 0.03689 & 
0.9629 & 0.03682  \cr 
      1915 & 520 & 3.683 & 0.2715 & 0.03016 & 0.03037 & 
0.9695 & 0.03031  \cr 
      1965 & 510 & 3.853 & 0.2595 & 0.02523 & 0.02542 & 
0.9744 & 0.02539  \cr 
      2015 & 495 & 4.071 & 0.2457 & 0.02009 & 0.02033 & 
0.9796 & 0.02032  \cr 
      2065 & 485 & 4.258 & 0.2349 & 0.01651 & 0.01670 & 
0.9832 & 0.01669  \cr 
      2115 & 470 & 4.500 & 0.2222 & 0.01281 & 0.01286 & 
0.9869 & 0.01285  \cr 
      2170 & 460 & 4.717 & 0.2120 & 0.01020 & 0.01022 & 
0.9895 & 0.01022  \cr 
      2225 & 450 & 4.944 & 0.2022 & 0.00805 & 0.00807 & 
0.9918 & 0.00807  \cr 
      2280 & 440 & 5.182 & 0.1930 & 0.00627 & 0.00634 & 
0.9936 & 0.00634  \cr 
      2340 & 425 & 5.506 & 0.1816 & 0.00447 & 0.00453 & 
0.9954 & 0.00453  \cr 
      2400 & 415 & 5.783 & 0.1729 & 0.00334 & 0.00340 & 
0.9966 & 0.00340  \cr 
      2460 & 405 & 6.074 & 0.1646 & 0.00247 & 0.00258 & 
0.9975 & 0.00258  \cr 
      2520 & 395 & 6.380 & 0.1567 & 0.00179 & 0.00190 & 
0.9982 & 0.00190  \cr 
      2585 & 385 & 6.714 & 0.1489 & 0.00126 & 0.00135 & 
0.9987 & 0.00135  \cr 
      2650 & 375 & 7.067 & 0.1415 & 0.00087 & 0.00093 & 
0.9991 & 0.00093  \cr 
      2720 & 370 & 7.351 & 0.1360 & 0.00065 & 0.00072 & 
0.9993 & 0.00072 \cr
\hfil&\hfil&\hfil&\hfil&\hfil&\hfil&\hfil&\hfil\cr}
\hrule}
\vfil
\endinsert

\subheading{\num{3.3.} Parallelograms}
 This second experiment investigates the hypothesis of 
conformal invariance
for simple curves $C$, namely parallelograms. The model is 
again $M_0$. 
One obvious
consequence of the conformal hypothesis is that the 
relative orientation of
the square lattice and of the parallelogram $C$ should be 
irrelevant in the measurement
of $\pi(E)$. This rotational symmetry is to be contrasted
with the obvious finite group of
symmetries of $M_0$.
Stronger consequences of full conformal
invariance can be tested by comparing the simulations
on a parallelogram that is not rectangular with 
the simulations on a rectangle.  

 Any parallelogram can
be obtained from the square $P_0$ by applying an element
$g$ of $GL(2,\Bbb R)$. If we take the square to be that 
defined
by the points $(0,0)$, $(1,0)$, $(1,1)$, $(0,1)$ and $g$
to be 
$$
  \left(\matrix a&b\\
                 c&d\endmatrix \right)
$$
then $P=gP_0$ is given by
$(0,0)$, $(a,c)$, $(a+b,c+d)$, $(b,d)$.
We stress that $g$ is not conformal.

 If, for example, $\pi_h(P,M)$ is the probability of a 
horizontal
crossing for large dilations of $P$ with respect to the
model $M$ then
$$
    \pi_h(P,M_0)=\pi_h(P_0,g^{-1}M_0)\tag 3.3a
$$
may be thought of as a coordinate of
the model $g^{-1}M_0$, that defined
by a horizontal crossing of dilations of $P_0$. With our 
identification
of the image of $\psi$ with the upper half-plane, the point 
$g^{-1}M_0$ corresponds to
$$
      \frac{ai+c}{bi+d}=\frac{a-ci}{b-di}.
$$   
If we identify in the natural way $\Bbb R^2$ with the 
complex numbers,
this is (unfortunately) the complex conjugate of the 
quotient
of the vertical by the horizontal side of $P$.

 To verify conformal invariance by simulation, we use the
conformal map $\varphi$ sending the unit circle to $P$,
and  $w_0$, whose value will depend
on $P$, to $(0,0)$, the point $-\bar w_0$ to $(a,c)$, the 
point $-w_0$
to  $(a+b,c+d)$,
and $\bar w_0$ to $(b,d)$. There will be exactly one
rectangle  $P_1$ with vertices $(0,0)$,
$(h,0)$, $(h,v)$, $(0,v)$ that is conformally equivalent to
$P$ together with its vertices, 
or to the unit circle together with the
specified four vertices. With $r=h/v$, conformal invariance
entails the relations
$$
  \pi_h(P,M_0)=\eta_h(r);\quad\pi_v(P,M_0)=
      \eta_v(r);\quad\pi_{hv}(P,M_0)=\eta_{hv}(r).
$$
Thus, in effect, given a parallelogram we find, in two steps
a conformal map that takes its interior to the interior
of a rectangle and takes vertices to vertices and
sides to sides.
Since the intermediate
curve is a circle with four distinguished points,
we have a choice. We can compare 
$\hat\pi_h(P,M_0)$ and $\hat\pi_v(P,M_0)$
with Cardy's predictions, or we can compare
them with  
the values for 
$\pi_h(P_1,M_0)$ and $\pi_v(P_1,M_0)$
interpolated from those given 
in Table 3.2 of the previous section. 
We prefer to compare with Cardy's predicted
values. 
For $\hat\pi_{hv}(P,M_0)$
however, we have only the second alternative. 

 The values of $\pi_d(P,M_0)$ can
also be predicted by Cardy's
formula, but only after they are precisely defined.
They can be defined\ as the probability of a crossing 
between
the upper half of the left side of $P$ to the right half of
the bottom side. If, on the other hand, $\alpha$ and 
$\beta$ are the
images of the upper half of the left side of $P_1$
and the right half of the bottom side, they can also
be defined as the probability of a crossing from $\alpha$ 
to 
$\beta$. Both definitions were used, according to the whim
of the individual experimenter, and we shall distinguish 
them as the first and
second definitions. 

 Although superfluous we provide in Figure 3.3 some curves
in the upper half-plane on which
conformal invariance implies that the 
three functions $\pi_h(P_0,M)$, 
$\pi_v(P_0,M)$, and $\pi_{hv}(P_0,M)$, taken as functions of
$z=\psi(M)$, are constant. 

\fighere{17 pc}\caption{Figure 3.3. \rm Two parallelograms 
with
vertices $(0,1, \tau_1+1,\tau_1)$ and $(0,1,\tau_2+
1,\tau_2)$
will have the same $\pi_h$ if and only if $\tau_1$ and 
$\tau_2$
lie on the same curve.}

 To obtain these curves we employ the Schwarz-Christoffel
transformation,
$$
\varphi: 
w\rightarrow\int_0^w(u^2-w_0^2)^{\alpha-1}(u^2-\bar 
w_0^2)^{-\alpha}\,du
=\frac1w\int_0^1(u^2-\epsilon_1^2)^{\alpha-1}(u^2-%
\epsilon_2^2)^{-\alpha}\,du,
$$
with
$$
    \epsilon_1=\frac{w_0}{w},\qquad\epsilon_2=\frac{\bar 
w_0}{w}.
$$
It maps the circle to a parallelogram with vertices,
in clockwise order, $\varphi(w_0)$,
$\varphi(\bar w_0)$, $\varphi(-w_0)$, $\varphi(-\bar w_0)$.
The interior angle at the vertex $\varphi(w_0)$
is $\alpha\pi$. It does not matter that the parallelogram 
is not in standard 
position.
  
 Fixing $w_0$ and letting $\alpha$ vary from $0$ to $1$, 
we obtain one of
the curves in Figure 3.3 as the collection of points
$$
     z= \frac{\varphi(w_0)-\varphi(\bar 
w_0)}{\varphi(w_0)-\varphi(-\bar w_0)}.
$$

 As parameters for a parallelogram, we can take $\alpha$ 
and $w_0$, or more 
conveniently $\alpha$ and the quotient $r$ of the lengths 
of the two sides.
To conform with the notation of \cite{U} we take 
$r=1/\vert z\vert$.
The data in Table 3.3 are from sixteen sets of 
experiments, corresponding to
four values of $\alpha$: $1/2$, $3/8$, $1/4$, $1/8$. 
In addition we chose four positions for the parallelogram,
one in which a side was parallel to the imaginary axis 
(labelled as
the case $\theta_0$), and then
rotations of this clockwise
through angles $\theta_1=\pi/12$, $\theta_2=\pi/6$, and
$\theta_3=\pi/4$. Conformal invariance entails, as 
observed, rotational
invariance, so that the rotation of the parallelogram
should not affect the result. In each experiment there 
were eleven
values for $r$, chosen so that the values of $\hat\pi_h$ 
were
about the same in each experiment and covered a 
representative
range.  The data are divided into four sets, each 
corresponding to 
a given value of $\alpha$. In each set the values of the 
various
crossing probabilities for different values of $\theta$ 
are listed
in adjacent columns to facilitate visual comparison. The 
probabilities $\pi_d$ are those given by the first 
definition.
The sample size was 100,000. 
The lengths of the sides were then chosen so that
there would be about $40000$ sites inside the parallelogram.
As we observed in \cite {U} and \S3.1, with this number
of sites an error
of about five parts in a thousand is to be expected. 
There appears to be
a systematic error of this order in the data.
For example the experimental values corresponding to
the true value $\pi^{cft}=.5$ are largely less than
$.5$. 
When the parallelogram
is not a rectangle with sides parallel to the axes, the 
collection
of sites within the parallelogram has an irregular 
boundary. We were
not able to find a method for accounting systematically for
the errors resulting from the anfractuosities. 

The measurements $\hat\pi_h,\hat\pi_v,\hat\pi_{hv}$ and 
$\hat\pi_d$ for all
values of the angle at the vertex $\alpha$, of the angle 
of rotation $\theta$
and of the ratio $r$ agree with the corresponding 
$\pi^{cft}(E)$ or $\hat\pi_{hv}^\square$
within the statistical errors and limitations due to the 
finiteness of the
lattices. Examining the rows at which $\pi^{cft}(E)=.5$,
we see that the worst discrepancies are $.0045$
for $\alpha=1/2$, $.0024$ for $\alpha=3/8$,
$.0039$ for $\alpha=1/4$, and $.0057$ for $\alpha=1/8$. 
As $\alpha$ grows smaller, the parallelogram grows
more skew, and the finite size of our lattices less and less
tolerable. For a given number of lattice points and
sufficiently small $\alpha$ the simulations
no longer make any sense, but $\alpha=1/8$ yields
acceptable results.

\pageinsert
\toptablecaption{\hsize=\vsize
\smc Table 3.3. \rm $\hat\pi_h,\hat\pi_v,\hat\pi_{hv}$ on 
parallelograms with angle
$\alpha\pi$ and with one side inclined at an angle 
$\theta_i$ to the
imaginary axis.}
\vbox{
\centerline{\hskip 5 cm$\alpha=1/2$}
\vskip 1 pc
\offinterlineskip 
\hrule height1.5pt
\halign{
\strut
\vrule width 1.5pt\quad\hfil#\hfil\quad&
\vrule width .5pt\quad\hfil#\hfil\quad&
\vrule width .5pt\quad\hfil#\hfil\quad&
\vrule width .5pt \quad\hfil#\hfil\quad&
\vrule width .5pt\quad\hfil#\hfil\quad&
\vrule width .5pt\quad\hfil#\hfil\quad&
\vrule width .5pt\quad\hfil#\hfil\quad&
\vrule width .5pt\quad\hfil#\hfil\quad&
\vrule width .5pt\quad\hfil#\hfil\quad&
\vrule width .5pt\quad\hfil#\hfil\quad&
\vrule width .5pt\quad\hfil#\hfil\quad
\vrule width 1.5pt \cr
ratio&
$\hat\pi_h(\theta_0)$&$\hat\pi_h(\theta_1)$&$\hat\pi_h(%
\theta_2)$&$\hat\pi_h(\theta_3)$&$\pi_h^{cft}$&
$\hat\pi_v(\theta_0)$&$\hat\pi_v(\theta_1)$&$\hat\pi_v(%
\theta_2)$&$\hat\pi_v(\theta_3)$&$\pi_v^{cft}$\cr
\noalign{\hrule height 1pt}
3.0000&.0627&.0609&.0608&.0619&.0617&
.9396&.9377&.9382&.9357&.9383\cr
2.3258&.1242&.1231&.1239&.1239&.1249&
.8759&.8752&.8739&.8730&.8751\cr
1.9041&.1973&.1927&.1920&.1922&.1943&
.8066&.8065&.8035&.7998&.8057\cr
1.4848&.3049&.3008&.2975&.3002&.3013&
.7018&.6974&.6972&.6978&.6987\cr
1.2198&.3984&.3978&.3951&.3968&.3977&
.6061&.6034&.5996&.5979&.6023\cr
1.0000&.5028&.4987&.4978&.4975&.5000&
.5008&.4999&.4974&.4955&.5000\cr
0.8198&.6020&.6039&.5996&.6030&.6023&
.3987&.3946&.3933&.3902&.3977\cr
0.6735&.7006&.6986&.6968&.7012&.6987&
.3006&.2977&.2986&.2952&.3013\cr
0.5252&.8074&.8050&.8078&.8057&.8057&
.1940&.1919&.1926&.1907&.1943\cr
0.4300&.8763&.8743&.8744&.8731&.8751&
.1253&.1235&.1220&.1227&.1249\cr
0.3333&.9388&.9373&.9388&.9385&.9383&
.0605&.0620&.0599&.0603&.0617\cr
\noalign{\hrule
\vskip 1.5 pt
\hrule }
ratio&
$\hat\pi_{hv}(\theta_0)$&$\hat\pi_{hv}(\theta_1)$&$\hat%
\pi_{hv}(\theta_2)$&$\hat\pi_{hv}(\theta_3)$&$\hat%
\pi_{hv}^{\square}$&
$\hat\pi_d(\theta_0)$&$\hat\pi_d(\theta_1)$&$\hat\pi_d(%
\theta_2)$&$\hat\pi_d(\theta_3)$&$\pi_d^{cft}$\cr
\noalign{\hrule height 1pt}
3.0000&.0623&.0605&.0605&.0615&.0616&
.1484&.1469&.1415&.1456&.1469\cr
2.3258&.1213&.1201&.1210&.1208&.1223&
.2044&.2048&.2035&.2037&.2055\cr
1.9041&.1861&.1818&.1813&.1805&.1837&
.2503&.2474&.2507&.2453&.2496\cr
1.4848&.2636&.2597&.2564&.2589&.2606&
.2934&.2947&.2906&.2947&.2942\cr
1.2198&.3069&.3061&.3016&.3027&.3057&
.3167&.3161&.3119&.3158&.3165\cr
1.0000&.3248&.3216&.3194&.3179&.3223&
.3200&.3232&.3228&.3185&.3244\cr
0.8198&.3061&.3035&.3025&.3004&.3057&
.3156&.3135&.3172&.3099&.3165\cr
0.6735&.2603&.2574&.2584&.2559&.2606&
.2944&.2938&.2922&.2893&.2942\cr
0.5252&.1836&1809&.1819&.1798&.1837&
.2500&.2491&.2470&.2498&.2496\cr
0.4300&.1223&.1206&.1190&.1201&.1223&
.2043&.2021&.2012&.2004&.2055\cr
0.3333&.0601&.0616&.0596&.0600&.0616&
.1463&.1471&.1466&.1432&.1469\cr}
\hrule height1.5pt
}
\vfil
\endinsert

\pageinsert
\toptablecaption{\hsize=\vsize
\smc Table 3.3. \rm{(continued)}}
\vbox{
\centerline{\hskip 5 cm$\alpha=3/8$}
\vskip 1 pc
\offinterlineskip 
\hrule height1.5pt
\halign{
\strut
\vrule width 1.5pt\quad\hfil#\hfil\quad&
\vrule width .5pt\quad\hfil#\hfil\quad&
\vrule width .5pt\quad\hfil#\hfil\quad&
\vrule width .5pt \quad\hfil#\hfil\quad&
\vrule width .5pt\quad\hfil#\hfil\quad&
\vrule width .5pt\quad\hfil#\hfil\quad&
\vrule width .5pt\quad\hfil#\hfil\quad&
\vrule width .5pt\quad\hfil#\hfil\quad&
\vrule width .5pt\quad\hfil#\hfil\quad&
\vrule width .5pt\quad\hfil#\hfil\quad&
\vrule width .5pt\quad\hfil#\hfil\quad
\vrule width 1.5pt \cr
ratio&
$\hat\pi_h(\theta_0)$&$\hat\pi_h(\theta_1)$&$\hat\pi_h(%
\theta_2)$&$\hat\pi_h(\theta_3)$&$\pi_h^{cft}$&
$\hat\pi_v(\theta_0)$&$\hat\pi_v(\theta_1)$&$\hat\pi_v(%
\theta_2)$&$\hat\pi_v(\theta_3)$&$\pi_v^{cft}$\cr
\noalign{\hrule height 1pt}
2.8661&.0615&.0611&.0603&.0600&.0608&
.9396&.9386&.9393&.9394&.9392\cr
2.2727&.1190&.1201&.1182&.1181&.1191&
.8789&.8824&.8798&.8814&.8809\cr
1.8428&.1920&.1929&.1907&.1911&.1939&
.8051&.8070&.8061&.8051&.8061\cr
1.4333&.3061&.3073&.3081&.3088&.3078&
.6939&.6929&.6919&.6921&.6922\cr
1.2092&.3955&.3945&.3950&.3918&.3962&
.6021&.6029&.6010&.6012&.6038\cr
1.0000&.5012&.4991&.5017&.4984&.5000&
.5008&.4986&.4976&.4982&.5000\cr
.8270&.6042&.6045&6035&.6041&.6038&
.3975&.3965&.3946&.3941&.3962\cr
.6977&.6910&.6928&.6924&.6920&.6922&
.3045&.3050&.3069&.3015&.3078\cr
.5427&.8091&.8062&.8072&.8056&.8061&
.1939&.1920&.1914&.1912&.1939\cr
.4400&.8809&.8829&.8810&.8821&.8809&
.1200&.1198&.1186&.1173&.1191\cr
.3489&.9394&.9383&.9377&.9381&.9392&
.0612&.0598&.0580&.0572&.0608\cr
\noalign{\hrule
\vskip 1.5 pt
\hrule }
ratio&
$\hat\pi_{hv}(\theta_0)$&$\hat\pi_{hv}(\theta_1)$&$\hat%
\pi_{hv}(\theta_2)$&$\hat\pi_{hv}(\theta_3)$&$\hat%
\pi_{hv}^{\square}$&
$\hat\pi_d(\theta_0)$&$\hat\pi_d(\theta_1)$&$\hat\pi_d(%
\theta_2)$&$\hat\pi_d(\theta_3)$&$\pi_d^{cft}$\cr
\noalign{\hrule}
2.8661&.0612&.0608&.0601&.0597&.0607&
.2046&.2023&.2038&.2041&.2048\cr
2.2727&.1164&.1177&.1156&.1155&.1169&
.2829&.2817&.2821&.2780&.2817\cr
1.8428&.1814&.1818&.1794&.1803&.1834&
.3477&.3484&.3490&.3491&.3487\cr
1.4333&.2633&.2636&.2638&.2639&.2645&
.4137&.4095&.4145&.4116&.4129\cr
1.2092&.3035&.3037&.3030&.3013&.3052&
.4346&.4343&.4396&.4399&.4402\cr
1.0000&.3231&.3224&.3225&.3204&.3223&
.4477&.4526&.4437&.4504&.4511\cr
.8270&.3064&.3046&.3044&.3039&.3052&
.4394&.4390&.4388&.4371&.4402\cr
.6977&.2615&.2611&.2636&.2597&.2645&
.4138&.4096&.4096&.4078&.4129\cr
.5427&.1835&.1809&.1805&.1799&.1834&
.3520&.3439&.3503&.3442&.3487\cr
.4400&.1175&.1171&.1162&.1151&.1169&
.2847&.2833&.2795&.2794&.2817\cr
.3489&.0609&.0594&.0576&.0569&.0607&
.2051&.2033&.2011&.1993&.2048\cr
    }
\hrule height1.5pt
}
\vfil
\endinsert

\pageinsert
\toptablecaption{\hsize=\vsize
{\smc Table 3.3.} {\rm(continued)}}
\vbox{
\centerline{\hskip 5 cm$\alpha=1/4$}
\vskip 1.5 pc
\offinterlineskip 
\hrule height1.5pt
\halign{
\strut
\vrule width 1.5pt\quad\hfil#\hfil\quad&
\vrule width .5pt\quad\hfil#\hfil\quad&
\vrule width .5pt\quad\hfil#\hfil\quad&
\vrule width .5pt \quad\hfil#\hfil\quad&
\vrule width .5pt\quad\hfil#\hfil\quad&
\vrule width .5pt\quad\hfil#\hfil\quad&
\vrule width .5pt\quad\hfil#\hfil\quad&
\vrule width .5pt\quad\hfil#\hfil\quad&
\vrule width .5pt\quad\hfil#\hfil\quad&
\vrule width .5pt\quad\hfil#\hfil\quad&
\vrule width .5pt\quad\hfil#\hfil\quad
\vrule width 1.5pt \cr
ratio&
$\hat\pi_h(\theta_0)$&$\hat\pi_h(\theta_1)$&$\hat\pi_h(%
\theta_2)$&$\hat\pi_h(\theta_3)$&$\pi_h^{cft}$&
$\hat\pi_v(\theta_0)$&$\hat\pi_v(\theta_1)$&$\hat\pi_v(%
\theta_2)$&$\hat\pi_v(\theta_3)$&$\pi_v^{cft}$\cr
\noalign{\hrule height 1pt}
2.3899&.0655&.0635&.0645&.0677&.0658&
.9331&.9340&.9345&.9329&.9342\cr
1.9885&.1189&.1177&.1177&.1216&.1191&
.8803&.8794&.8812&.8782&.8809\cr
1.6354&.1985&.1959&.2016&.2050&.2006&
.7971&.7971&.8003&.7968&.7994\cr
1.3443&.3073&.3041&.3059&.3090&.3072&
.6900&.6912&.6935&.6885&.6928\cr
1.1674&.3965&.3926&.3939&.3972&.3961&
.6007&.6019&.6017&.6009&.6039\cr
1.0000&.4971&.4961&.5000&.5031&.5000&
.4970&.4994&.5007&.4963&.5000\cr
.8566&.6046&.6033&.6045&.6059&.6039&
.3920&.3957&.3935&.3924&.3961\cr
.7439&.6889&.6913&.6923&.6941&.6928&
.3059&.3050&.3040&.3030&.3072\cr
.6115&.7971&.7971&.7986&.8020&.7994&
.1998&.1997&.2019&.1988&.2006\cr
.5029&.8803&.8786&.8811&.8807&.8809&
.1210&.1174&.1182&.1182&.1191\cr
.4184&.9342&.9336&.9356&.9342&.9342&
.0636&.0653&.0638&.0655&.0658\cr
\noalign{\hrule
\vskip 1.5 pt
\hrule }
ratio&
$\hat\pi_{hv}(\theta_0)$&$\hat\pi_{hv}(\theta_1)$&$\hat%
\pi_{hv}(\theta_2)$&$\hat\pi_{hv}(\theta_3)$&$\hat%
\pi_{hv}^{\square}$&
$\hat\pi_d(\theta_0)$&$\hat\pi_d(\theta_1)$&$\hat\pi_d(%
\theta_2)$&$\hat\pi_d(\theta_3)$&$\pi_d^{cft}$\cr
\noalign{\hrule}
 2.3899&.0651&.0631&.0640&.0671&.0656&
.3112&.3022&.3054&.3106&.3089\cr
1.9885&.1163&.1154&.1150&.1189&.1169&
.4049&.3973&.4035&.4061&.4044\cr
1.6354&.1859&.1847&.1898&.1922&.1890&
.4953&.4958&.4973&.5037&.4984\cr
1.3443&.2630&.2608&.2637&.2637&.2641&
.5666&.5706&.5710&.5712&.5707\cr
1.1674&.3037&.3023&.3016&.3048&.3052&
.5994&.5961&.5994&.6008&.6026\cr
1.0000&.3190&.3206&.3216&.3212&.3223&
.6116&.6089&.6105&.6134&.6150\cr
.8566&.3024&.3051&.3021&.3033&.3052&
.5988&.5968&.6015&.5961&.6026\cr
.7439&.2628&.2615&.2607&.2610&.2641&
.5674&.5702&.5687&.5657&.5707\cr
.6115&.1877&.1875&.1895&.1867&.1890&
.4962&.4953&.4993&.4949&.4984\cr
.5029&.1184&.1147&.1155&.1156&.1169&
.4016&.4013&.4009&.4017&.4044\cr
.4184&.0633&.0649&.0634&.0648&.0656&
.3049&.3025&.3067&.3074&.3089\cr
   }
\hrule height1.5pt
}
\vfil
\endinsert

\pageinsert
\toptablecaption{\hsize=\vsize
{\smc Table 3.3.} {\rm(continued)}}
\vbox{
\centerline{\hskip 5 cm$\alpha=1/8$}
\vskip 1.5 pc
\offinterlineskip 
\hrule height1.5pt
\halign{
\strut
\vrule width 1.5pt\quad\hfil#\hfil\quad&
\vrule width .5pt\quad\hfil#\hfil\quad&
\vrule width .5pt\quad\hfil#\hfil\quad&
\vrule width .5pt \quad\hfil#\hfil\quad&
\vrule width .5pt\quad\hfil#\hfil\quad&
\vrule width .5pt\quad\hfil#\hfil\quad&
\vrule width .5pt\quad\hfil#\hfil\quad&
\vrule width .5pt\quad\hfil#\hfil\quad&
\vrule width .5pt\quad\hfil#\hfil\quad&
\vrule width .5pt\quad\hfil#\hfil\quad&
\vrule width .5pt\quad\hfil#\hfil\quad
\vrule width 1.5pt \cr
ratio&
$\hat\pi_h(\theta_0)$&$\hat\pi_h(\theta_1)$&$\hat\pi_h(%
\theta_2)$&$\hat\pi_h(\theta_3)$&$\pi_h^{cft}$&
$\hat\pi_v(\theta_0)$&$\hat\pi_v(\theta_1)$&$\hat\pi_v(%
\theta_2)$&$\hat\pi_v(\theta_3)$&$\pi_v^{cft}$\cr
\noalign{\hrule height 1pt}
1.7926&.0605&.0607&.0582&.0610&.0611&
.9378&.9374&.9383&.9373&.9389\cr
1.5342&.1221&.1231&.1224&.1213&.1238&
.8747&.8754&.8763&.8744&.8762\cr
1.3429&.2087&.2078&.2065&.2062&.2081&
.7878&.7892&.7901&.7876&.7919\cr
1.2097&.2999&.2929&.2971&.2947&.2971&
6998.&.7035&.7014&.7004&.7029\cr
1.1047&.3853&.3870&.3852&.3851&.3893&
.6088&.6099&.6065&.6087&.6107\cr
1.0000&.4967&.5002&.4984&.4943&.5000&
.4987&.4990&.4969&.4966&.5000\cr
0.9053&.6080&.6084&.6079&.6064&.6107&
.3851&.3895&.3875&.3855&.3893\cr
0.8266&.7005&.6995&.7006&.6981&.7029&
.2943&.2955&.2970&.2909&.2971\cr
0.7446&.7878&.7885&.7893&.7860&.7919&
.2097&.2087&.2098&.2049&.2081\cr
0.6518&.8765&.8761&.8756&.8735&.8762&
.1235&.1229&.1240&.1201&.1238\cr
0.5579&.9404&.9391&.9391&.9372&.9389&
.0606&.0613&.0605&.0597&.0611\cr
\noalign{\hrule
\vskip 1.5 pt
\hrule }
ratio&
$\hat\pi_{hv}(\theta_0)$&$\hat\pi_{hv}(\theta_1)$&$\hat%
\pi_{hv}(\theta_2)$&$\hat\pi_{hv}(\theta_3)$&$\hat%
\pi_{hv}^{\square}$&
$\hat\pi_d(\theta_0)$&$\hat\pi_d(\theta_1)$&$\hat\pi_d(%
\theta_2)$&$\hat\pi_d(\theta_3)$&$\pi_d^{cft}$\cr
\noalign{\hrule height 1pt}
1.7926&.0601&.0604&.0578&.0608&.0610&
.5687&.5691&.5668&.5725&.5708\cr
1.5342&.1193&.1202&.1199&.1185&.1213&
.7020&.7034&.6987&.7038&.7021\cr
1.3429&.1945&.1939&.1932&.1928&.1952&
.7878&.7801&.7804&.7772&.7819\cr
1.2097&.2593&.2543&.2577&.2553&.2581&
.8273&.8202&.8239&.8192&.8236\cr
1.1047&.2999&.3013&.2994&.2997&.3028&
.8470&.8450&.8414&.8415&.8452\cr
1.0000&.3203&.3218&.3195&.3180&.3223&
.8530&.8522&.8518&.8507&.8533\cr
.9053&.2987&.3006&.2999&.2984&.3028&
.8456&.8451&.8424&.8441&.8452\cr
.8266&.2558&.2558&.2568&.2514&.2581&
.8228&.8226&.8214&.8193&.8236\cr
.7446&.1957&.1941&.1957&.1917&.1952&
.7834&.7805&.7810&.7801&.7819\cr
.6518&.1205&.1203&.1211&.1175&.1213&
.6996&.7039&.7029&.6957&.7021\cr
.5579&.0602&.0611&.0601&.0593&.0610&
.5654&.5695&.5666&.5666&.5708\cr
    }
\hrule height1.5pt
}
\vfil
\endinsert                   
                    
\subheading{\num{3.4.} Striated models} 
 By numerical simulation we showed in \cite{U} that the 
four functions
$\pi_h$, $\pi_v$, $\pi_{hv}$ and $\pi_d$ coincided for the 
six following
models: percolation by sites and by bonds on square, 
triangular and hexagonal
lattices. This gives some support to the hypothesis of 
universality. 
Because of the symmetry of these regular lattices, 
the matrix $g$ predicted by the hypothesis is
diagonal in all six 
cases. This need not to be so, as the following example 
shows.

 If we restrict ourselves to percolation by sites, it is 
pretty 
clear that, within the limits of experimental observation,
the most general case can be obtained by choosing
on the lattice $L=\Bbb Z^2$ probabilities $p(s)$ that depend
on the position of $s$ modulo some sub-lattice $NL$, where 
$N$ is a very
large integer. This is certainly convenient
for simulations. In particular to obtain a model that does
not yield a point on the imaginary axis, we can deliberately
skew the usual model by insisting that the probabilities 
be close
to $0$ along some band athwart the lattice and otherwise,
as far as conditions of periodicity permit, close to $1$.
This we call a striated model. The hypothesis of 
universality
implies that any model $M$, and in particular any striated
model, corresponds to a point in the upper half-plane, and 
that, once
this point is known, all probabilities $\pi(E,M)$ can be 
calculated
from those for percolation by sites on a square
lattice. Since in the eyes of many of our colleagues, 
universality
even in the form proposed in the hypothesis
is a commonplace, widely\ accepted and well understood, we 
have confined ourselves here to
the examination of a single example.
It illustrates adequately the hypothesis, and the 
calculation
of 
an approximation to the associated matrix $g$ is a useful 
exercise.

 The band is constructed
by periodicity from a rectangle 
with $6\times 4$ sites, as in Figure 3.4. The
points on the band are $(0,0)$,
$(1,0)$, $(1,1)$, $(2,1)$,
$(3,2)$, $(4,2)$,
$(4,3)$, and $(5,3)$.
Thus $N=12$. (The bands are depicted by black squares
in Figure 3.4.)
All other points are off the bands.
The probability $p_1$ on the bands
is one-fifth the probability $p_2$ off the
bands. Simulation and the technique of
\cite{U} yield a value $p_2=0.84928$
for the critical probability.
The band forms an angle 
whose tangent is $2/3$ with the x-axis and
we can expect that the model corresponds to a dilation of 
one axis
of the model $M_0$ followed by a rotation of
approximately this angle, for what we have done is
to hinder percolation perpendicular to the band,
and to foster it parallel to the
band. Thus the model presumably behaves like
site-percolation on a rectangular lattice in which the basic
rectangle has its long side parallel to the band.

\fighere{11 pc}\caption{Figure 3.4. \rm Definition of a 
striated
model. Black sites (tiles) are open with probability $p_1$ 
and 
white ones with probability $p_2=5p_1$. }

 According to the hypothesis of universality
there will be a matrix $g$ such that
$$
    \pi(E,M)=\pi(E,gM_0),\tag 3.4a
$$
for all events $E$. To calculate an approximation to
$g$ we consider first the
events defined by a horizontal crossing
of a rectangle  $R_r$ of aspect ratio $r$ with sides
parallel to the coordinate axes.
It is clear from our discussion of conformal invariance
for parallelograms that,
modulo the group of linear conformal transformations
acting on the right and the group
$$
\left\{\left(\matrix 
\pm1&0\\0&\pm1\endmatrix\right)\right\} \tag 3.4b
$$
acting on the left, there is at most
one matrix $g$ such that
$$
    \pi_h(R_r,M)=\pi_h(R_r,gM_0)=\pi_f(g^{-1}R_r,M_0)\tag 
3.4c
$$
for all $r$, or even for two values of $r$.
We repeat that the second equality is formal.
We take, to be precise, $g^{-1}$ in the form
$$
  \left(\matrix a\sin\theta&0\\
                -a\cos\theta&1\endmatrix \right),
$$
with $0\leq \theta\leq \pi$. The angle $\theta$
is the interior angle of the parallelograms $g^{-1}R_r$
and the right side of (3.4c) is calculated
from Cardy's formula
by the methods described in \S 3.3. The matrix $g$ itself is
then a scalar multiple of
$$
  \left(\matrix 1&0\\
                a\cos\theta&a\sin\theta\endmatrix \right).
$$
Equally useful is the relation 
$$
    \pi_v(R_r,M)=\pi_v(R_r,gM_0)=\pi_v(g^{-1}R_r,M_0).\tag 
3.4d 
$$

 Universality affirms that for the given striated
model $M$ a matrix $g$ can be found
such that
(3.4a) is satisfied for
all events $E$. The equations (3.4c)
and (3.4d) are particular cases of
(3.4a) that almost suffice to determine $g$.
In Table 3.4a we give the left side of (3.4c) and (3.4d) 
for 41 values of
$r$, or rather values
obtained for the left side
by simulation. The method of least squares was then used 
to find
values of $a$ and $\theta$ that minimized the difference 
between the two
sides of (3.4c) or (3.4d), the right side
being determined
as described in \S 3.3. The values obtained are:
$$
    \hat a=0.7538\qquad\qquad\hat\theta=0.2643\pi.
$$
Let $\hat g$ be the associated matrix.
For each value of $r$, the two parameters  $\hat a$ and 
$\hat\theta$
are used to calculate,
from Cardy's formula,
the aspect ratios $r_0$ of the 
rectangles such that
the numbers
$\pi_h^{cft}$
and $\pi_v^{cft}$
appearing in
the row of Table 3.4a labeled by $r$
are hypothetically equal
by universality (assuming $\hat g$ is the matrix
appearing in (3.4a)) to
$\pi_h(R_{r_0}, M_0)$ and $\pi_v(R_{r_0}, M_0)$.

\pageinsert
\toptablecaption{
\smc Table 3.4\rm a. Data for calculating
the matrix $\hat g$ of the striated model.}
\centerline{\vbox{\offinterlineskip
\hrule
\halign{\strut\vrule\quad\hfil#\hfil\quad&\vrule\quad%
\hfil#\hfil\quad&    \vrule\quad\hfil#\hfil\quad&\vrule%
\quad\hfil#\hfil\quad&
    \vrule\quad\hfil#\hfil\quad&\vrule\quad\hfil#\hfil\quad
    \vrule\cr
\hfil&\hfil&\hfil&\hfil&\hfil&\hfil\cr
\hfil $r$\hfil& \hfil$r_0$\hfil& 
\hfil$\hat\pi_h$\hfil&\hfil 
$\pi_h^{cft}$\hfil&\hfil$\hat\pi_v$\hfil&\hfil
$\pi_{v}^{cft}$\hfil\cr
\hfil&\hfil&\hfil&\hfil&\hfil&\hfil\cr
\noalign{\hrule height1pt}
\hfil&\hfil&\hfil&\hfil&\hfil&\hfil\cr
     0.6070 & 0.3873 & 0.9058 & 0.9045 & 0.0965 & 0.0955 
\cr 
     0.6400 & 0.4116 & 0.8885 & 0.8880 & 0.1146 & 0.1120 
\cr 
     0.6721 & 0.4356 & 0.8716 & 0.8711 & 0.1302 & 0.1289 
\cr 
     0.7059 & 0.4613 & 0.8546 & 0.8527 & 0.1492 & 0.1473 
\cr 
     0.7414 & 0.4887 & 0.8344 & 0.8327 & 0.1699 & 0.1673 
\cr 
     0.7753 & 0.5153 & 0.8147 & 0.8131 & 0.1881 & 0.1869 
\cr 
     0.8190 & 0.5502 & 0.7891 & 0.7874 & 0.2148 & 0.2126 
\cr 
     0.8611 & 0.5845 & 0.7641 & 0.7623 & 0.2388 & 0.2377 
\cr 
     0.9048 & 0.6206 & 0.7378 & 0.7361 & 0.2672 & 0.2639 
\cr 
     0.9512 & 0.6599 & 0.7114 & 0.7083 & 0.2933 & 0.2917 
\cr 
     1.000 & 0.7018 & 0.6801 & 0.6793 & 0.3228 & 0.3207 \cr 
     1.051 & 0.7467 & 0.6521 & 0.6492 & 0.3534 & 0.3508 \cr 
     1.105 & 0.7948 & 0.6210 & 0.6181 & 0.3832 & 0.3819 \cr 
     1.161 & 0.8457 & 0.5893 & 0.5867 & 0.4145 & 0.4133 \cr 
     1.221 & 0.9007 & 0.5562 & 0.5543 & 0.4458 & 0.4457 \cr 
     1.290 & 0.9651 & 0.5188 & 0.5185 & 0.4816 & 0.4815 \cr 
     1.349 & 1.021 & 0.4909 & 0.4891 & 0.5133 & 0.5109 \cr 
     1.417 & 1.086 & 0.4594 & 0.4570 & 0.5455 & 0.5430 \cr 
     1.488 & 1.155 & 0.4271 & 0.4252 & 0.5770 & 0.5748 \cr 
     1.562 & 1.229 & 0.3957 & 0.3938 & 0.6086 & 0.6062 \cr 
     1.647 & 1.313 & 0.3606 & 0.3607 & 0.6396 & 0.6393 \cr 
     1.730 & 1.395 & 0.3302 & 0.3309 & 0.6692 & 0.6691 \cr 
     1.824 & 1.490 & 0.3003 & 0.2998 & 0.7008 & 0.7002 \cr 
     1.910 & 1.576 & 0.2750 & 0.2738 & 0.7277 & 0.7262 \cr 
     2.014 & 1.681 & 0.2463 & 0.2453 & 0.7546 & 0.7547 \cr 
     2.124 & 1.792 & 0.2204 & 0.2183 & 0.7836 & 0.7817 \cr 
     2.224 & 1.894 & 0.1961 & 0.1963 & 0.8059 & 0.8037 \cr 
     2.336 & 2.008 & 0.1758 & 0.1742 & 0.8277 & 0.8258 \cr 
     2.453 & 2.127 & 0.1538 & 0.1538 & 0.8477 & 0.8462 \cr 
     2.597 & 2.274 & 0.1326 & 0.1319 & 0.8695 & 0.8681 \cr 
     2.727 & 2.407 & 0.1159 & 0.1147 & 0.8855 & 0.8853 \cr 
     2.864 & 2.547 & 0.0990 & 0.0991 & 0.9010 & 0.9009 \cr 
     3.017 & 2.703 & 0.0846 & 0.0842 & 0.9158 & 0.9159 \cr 
     3.142 & 2.830 & 0.0744 & 0.0737 & 0.9269 & 0.9263 \cr 
     3.309 & 3.001 & 0.0618 & 0.0616 & 0.9396 & 0.9384 \cr 
     3.495 & 3.191 & 0.0512 & 0.0505 & 0.9497 & 0.9495 \cr 
     3.683 & 3.382 & 0.0410 & 0.0413 & 0.9590 & 0.9587 \cr 
     3.853 & 3.556 & 0.0346 & 0.0344 & 0.9661 & 0.9656 \cr 
     4.071 & 3.778 & 0.0279 & 0.0273 & 0.9734 & 0.9727 \cr 
     4.258 & 3.969 & 0.0230 & 0.0223 & 0.9780 & 0.9777 \cr 
     4.500 & 4.217 & 0.0174 & 0.0172 & 0.9830 & 0.9828 \cr
\hfil&\hfil&\hfil&\hfil&\hfil&\hfil\cr}
\hrule}
}
\vfil
\endinsert

 The ambiguity
entailed by multiplication
by the matrices (3.4b) implies
that the value $\hat\theta=\pi-0.2643\pi$ is also possible;
it leads to the same values of the right sides.
Thus a second experiment is required to eliminate it.

 Once estimates for $a$ and $\theta$ have been obtained, 
then for any parallelogram $P$
predicted
values of $\pi_h(P,M)$, $\pi_v(P,M)$,
and of $\pi_{hv}(P,M)$ can be calculated 
from the right side of (3.4a) and Cardy's formula or
by interpolation from Table 3.2, as in the
section on parallelograms.
As a first choice we took $P=\hat gR_{r_0}$,
because, for example, we expect that
$$
   \pi_h(gR_{r_0},M)=\pi_h(R_{r_0},M_0)=\eta_h(r_0).
$$
One interior angle of the parallelogram $\hat gR_{r_0}$
would then be equal to $0.3502\pi$.

 The results appear in Table 3.4b, in which
the variable $r_0$ is the free
variable. Thus the
coordinates of the vertices of the 
parallelograms on the striated lattice actually used
are calculated from it. They are
$(0,0)$, $(0,b)$, $(c,d)$ and $(c,b+d)$, where the
integers $b$, $c$, $d$ assume the values in the table.
The values of $r_0$ are given in the table; the ratio of the
sides of the parallelogram $gR_{r_0}$ are then $r=\hat Br_0$
with
$$
     \hat B=\frac{\sqrt{(1+\hat 
a^2\cos^2\hat\theta)}}{\hat a\sin\hat\theta}=2.016.
$$
It is clear from this table that of the two possibilities
for $g$ modulo the group (3.4b) we have chosen the correct 
one,
for otherwise there would be no agreement between the
values obtained by simulation and the predicted values.

\pageinsert
\toptablecaption{\hsize=\vsize
\smc Table 3.4\rm b. $\hat\pi_h,\hat\pi_v,\hat\pi_{hv}$ 
for a
parallelogram $P=\hat gR_{r_0}$ on the striated model.}
\vbox{\offinterlineskip
\hrule
\halign{\strut\vrule\quad\hfil#\hfil
    \quad&\vrule\quad\hfil#\hfil\quad&
    \vrule\quad\hfil#\hfil\quad&\vrule\quad\hfil#\hfil\quad&
    \vrule\quad\hfil#\hfil\quad&\vrule\quad\hfil#\hfil\quad&
    \vrule\quad\hfil#\hfil\quad&\vrule\quad\hfil#\hfil\quad&
    \vrule\quad\hfil#\hfil\quad&\vrule\quad\hfil#\hfil\quad
    \vrule\cr
\hfil&\hfil&\hfil&\hfil&\hfil&\hfil&\hfil&\hfil&\hfil&%
\hfil\cr
\hfil $b$\hfil& \hfil$c$\hfil&\hfil $d$\hfil&
\hfil $r_0$\hfil& \hfil$\hat\pi_h$\hfil&\hfil 
$\pi_h^{cft}$\hfil&\hfil$\hat\pi_v$\hfil&\hfil
$\pi_{v}^{cft}$\hfil&
\hfil$\hat\pi_{hv}$\hfil&\hfil$\hat\pi_{hv}^{\square}$%
\hfil\cr
\hfil&\hfil&\hfil&\hfil&\hfil&\hfil&\hfil&\hfil&\hfil&%
\hfil\cr
\noalign{\hrule height1pt}
\hfil&\hfil&\hfil&\hfil&\hfil&\hfil&\hfil&\hfil&\hfil&%
\hfil\cr
300 & 539 &  274 & 1.000  & 0.5039 & 0.5000 & 0.4983 & 
0.5000 & 0.3229 & 0.3223 \cr 
390 & 736 &  374 & 1.050  & 0.4772 & 0.4746 & 0.5209 & 
0.5254 & 0.3195 & 0.3211 \cr 
380 & 754 &  383 & 1.104  & 0.4537 & 0.4487 & 0.5498 & 
0.5513 & 0.3195 & 0.3180 \cr 
372 & 776 &  394 & 1.160  & 0.4254 & 0.4229 & 0.5751 & 
0.5771 & 0.3133 & 0.3128 \cr 
362 & 794 &  403 & 1.220  & 0.3989 & 0.3974 & 0.5977 & 
0.6026 & 0.3044 & 0.3056 \cr 
352 & 815 &  414 & 1.288  & 0.3726 & 0.3701 & 0.6259 & 
0.6299 & 0.2961 & 0.2953 \cr 
344 & 833 &  424 & 1.348  & 0.3503 & 0.3477 & 0.6461 & 
0.6523 & 0.2855 & 0.2856 \cr 
336 & 855 &  435 & 1.416  & 0.3259 & 0.3237 & 0.6726 & 
0.6763 & 0.2734 & 0.2734 \cr 
328 & 877 &  446 & 1.488  & 0.3015 & 0.3003 & 0.6948 & 
0.6997 & 0.2591 & 0.2600 \cr 
320 & 898 &  456 & 1.561  & 0.2788 & 0.2781 & 0.7173 & 
0.7219 & 0.2453 & 0.2461 \cr 
312 & 923 &  469 & 1.646  & 0.2581 & 0.2545 & 0.7427 & 
0.7455 & 0.2320 & 0.2299 \cr
\hfil&\hfil&\hfil&\hfil&\hfil&\hfil&\hfil&\hfil&\hfil&%
\hfil\cr}
\hrule}
\vskip2pc
\toptablecaption{\hsize=\vsize
\smc Table 3.4\rm c. $\hat\pi_h,\hat\pi_v,\hat\pi_{hv}$
for a parallelogram with interior angle $3\pi/8$ on the 
striated model.}
\vbox{\offinterlineskip
\hrule
\halign{\strut\vrule\quad\hfil#\hfil
    \quad&\vrule\quad\hfil#\hfil\quad&
    \vrule\quad\hfil#\hfil\quad&\vrule\quad\hfil#\hfil\quad&
    \vrule\quad\hfil#\hfil\quad&\vrule\quad\hfil#\hfil\quad&
    \vrule\quad\hfil#\hfil\quad&
    \vrule\quad\hfil#\hfil\quad&\vrule\quad\hfil#\hfil\quad&
    \vrule\quad\hfil#\hfil\quad&\vrule\quad\hfil#\hfil\quad
    \vrule\cr
\hfil&\hfil&\hfil&\hfil&\hfil&\hfil&\hfil&\hfil&\hfil&%
\hfil&\hfil\cr
\hfil $b$\hfil& \hfil$c$\hfil&\hfil $d$\hfil&
\hfil $r$\hfil&\hfil $r_0$\hfil& 
\hfil$\hat\pi_h$\hfil&\hfil 
$\pi_h^{cft}$\hfil&\hfil$\hat\pi_v$\hfil&\hfil
$\pi_{v}^{cft}$\hfil&
\hfil$\hat\pi_{hv}$\hfil&\hfil$\hat\pi_{hv}^{\square}$%
\hfil\cr
\hfil&\hfil&\hfil&\hfil&\hfil&\hfil&\hfil&\hfil&\hfil&%
\hfil&\hfil\cr
\noalign{\hrule height1pt}
\hfil&\hfil&\hfil&\hfil&\hfil&\hfil&\hfil&\hfil&\hfil&%
\hfil&\hfil\cr
 362 & 601 & 444 &    1.000 & 1.000 & 0.5022 & 0.5000 & 
0.4956 & 0.5000 & 0.3211 & 0.3223 \cr 
 344 & 630 & 466 &    1.105 & 1.111 & 0.4474 & 0.4452 & 
0.5508 & 0.5548 & 0.3162 & 0.3174 \cr 
 329 & 660 & 488 &    1.210 & 1.224 & 0.3985 & 0.3959 & 
0.5999 & 0.6041 & 0.3048 & 0.3051 \cr 
 312 & 695 & 514 &    1.343 & 1.367 & 0.3440 & 0.3409 & 
0.6527 & 0.6591 & 0.2827 & 0.2822 \cr 
 292 & 743 & 549 &    1.534 & 1.573 & 0.2774 & 0.2747 & 
0.7212 & 0.7253 & 0.2451 & 0.2401 \cr 
 270 & 803 & 593 &    1.793 & 1.852 & 0.2069 & 0.2051 & 
0.7909 & 0.7949 & 0.1933 & 0.1927 \cr
\hfil&\hfil&\hfil&\hfil&\hfil&\hfil&\hfil&\hfil&\hfil&%
\hfil&\hfil\cr}
\hrule}
\vfil
\endinsert

 As a further verification we examined the probabilities
like $\pi_h(\hat gP,M)$
for a parallelogram $P$ of interior angle $3\pi/8$,
and with one pair of opposite sides vertical.
One
interior angle of the parallelogram $\hat gP$
is then very close to 
$0.2974\pi$. The values $r$ in Table 3.4c are the ratios
of sides of $\hat gP$. The values $r_0$ are the aspect 
ratios
of rectangles conformally equivalent to $P$, and are used to
calculate the predicted values given in Table 3.4c.

As for the previous experiment with parallelograms, a 
systematic
error can be seen: for example, in both Table 3.4b and 3.4c 
the value $\hat\pi_h$ 
is always larger than $\pi_h^{cft}$. Still the discrepancy 
is in the
third significant digit and comparable to the error due to 
the
finiteness of the lattice (see \S 3.1);
so the agreement is satisfactory. 
The only differences greater than
$.005$ are those at row $(344,833,424)$ in Table 3.4b
and at row $(312,695,514)$
in Table 3.4c, where
differences of $.0058$ and $.0064$ are found. The 
anfractuosities
at the boundary of $P$ may again have played a role.

\subheading{\num{3.5.} Exterior domains}
 Once the notion of conformal invariance has appeared
in a convincing manner
in the study of percolation in simply-connected bounded
planar regions, many other questions 
arise. First of all there is no reason
to confine oneself to simply connected regions,
nor, apart from experimental inconvenience, to 
bounded regions. Even the notion of crossing probability
can be considerably extended. 

 Limitations on memory force the simulations to be confined
to a bounded region, and
when examining unbounded regions it is necessary either to 
devise
an experiment that is not sensitive to the
inevitable hole at infinity, or
to estimate the error it causes. Moreover the boundaries 
of unbounded
regions, such
as the exterior of a
convex polygon, usually have angles that 
penetrate the region,
and these are the source of substantial errors in the 
simulation.

 Examining percolation by bonds on a square lattice, we
saw in \cite {U} that an indeterminacy of the order of the 
lattice mesh
led to an indeterminacy of about $1/d$ 
in the crossing probabilities, if $d$
is the diameter of the finite lattice.
This is to be expected by Cardy's
formula, at least for $\pi_h$. A modification of the order 
of $1$
in the endpoint  $z_i$ of $\alpha$ or $\beta$ entails a 
change in
$w_i$ and thus in the cross-ratio of about $1/d$. If, 
however,
$z_i$ were the vertex of a penetrating wedge
with exterior angle $\alpha\pi$, then near $w_i$ 
the function $\varphi$ behaves like
$(w-w_i)^\alpha$ and its inverse like $(z-z_i)^{1/\alpha}$.
Consequently, if for example $\alpha=1.5$,
an indeterminacy of say $.01$ is magnified to
one of $.01^{2/3}\sim.05$, and the data cease to be
persuasive.

 In order to avoid problems with penetrating angles, we 
have 
confined ourselves to experiments with circles. The 
obvious question
is whether percolation in the interior of the circle is 
equivalent
to percolation in the exterior, thus whether crossing 
probabilities are
invariant under the map $z\rightarrow1/z$.  Since this takes
the bounded domain $\vert z\vert \leq1$ to the unbounded 
domain
$\vert z\vert \geq 1$,  we are immediately confronted with 
the
impossibility of treating all lattice points in the 
exterior domain.

  Take a circle $C$ of radius $1$ centred at the origin, 
and let $\alpha$
be the arc of $C$ from $\tfrac{3\pi}{4}$
to $\tfrac{5\pi}{4}$ and $\beta$ its reflection in the
axis of ordinates. Conformal invariance implies that the 
probability
of a crossing from $\alpha'=A\alpha$ to $\beta'=A\beta$ in 
the exterior
of $C'=AC$ should be close to $.5$ for $A$ large.
Experiments can, however, only be carried out on
finite lattices. We can take, for example, percolation
inside the annulus formed by two circles, the inner
one having radius $A$ and the outer a radius
as large as time and the machines available allow,
and estimate the probability within this annulus of
a crossing from $\alpha'$ to $\beta'$. The results are 
disappointing.
For an inner radius of $100 $ and an outer
radius of $1000$ the probability is
about $.431$. With the same outer
radius and inner radii of $50$ and
$25$ the probabilities become
about $.457$ and $.468$, in every case
far short of the expected $.5$,
although the value is seen to improve
with increasing ratio of the two radii.
It is also clear, however, that
to achieve an
adequate value of the ratio and of the inner
radius would put impossible demands on machine
memory. Therefore it is necessary 
either to exploit 
methods of extrapolation 
or to devise other experiments
to test conformal invariance under inversion.

\fighere{9pc}\caption{Figure 3.5\rm a. Possible crossings 
for
$\pi_{h}^{int}$ and $\pi_{h}^{ext}$.}

 The most direct is to take the two concentric circles of 
radii
$r_1<r_2$ and to divide each into four arcs of equal length
symmetric about the axes. 
We consider only crossings within 
dilations of the annulus, and  we introduce the probability
$\pi_h^{int}$ of a crossing from  the left interior arc to 
the right
interior arc, as well as the probability
$\pi_h^{ext}$ of a crossing from the
left exterior arc to the right exterior arc. (See Figure 
3.5a.)
The two probabilities  $\pi_v ^{int}$ and  $\pi_v^{ext}$
are defined similarly. Conformal invariance under 
$z\rightarrow 1/z$
implies that all four are equal in the limit of large $r_1$
and fixed $r_2/r_1$. We also introduce $\pi_{hv}^{int}$
and $\pi_{hv}^{ext}$.

\midinsert
\toptablecaption{Table 3.5. \rm $\hat\pi_h,\hat\pi_v,
\hat\pi_{hv}$ for an annulus and a cylinder.}
\centerline{%
\vbox{
\offinterlineskip
\hrule height 1 pt
\halign{
&\vrule#&\strut\quad\hfil#\hfil\quad\cr
width 1 pt&\omit&&$\hat\pi_h$&&$\hat\pi_v$&&
$\hat\pi_{hv}$&width 1 pt\cr
width 1pt height 2 pt &\omit&width .5pt&\omit&width .5pt
&\omit&width .5pt&\omit&width 1pt\cr
\noalign{\hrule height 1 pt}
width 1pt height 1pt &\omit&width .5pt&\omit&width .5pt
&\omit&width .5pt&\omit&width 1pt\cr
width1pt&interior&& .4316&& .4306&& .2539&width1pt\cr
\noalign{\hrule}
width 1pt height 1pt &\omit&width .5pt&\omit&width .5pt
&\omit&width .5pt&\omit&width 1pt\cr
width1pt&exterior&& .4356&& .4348&& .2586&width1pt\cr
\noalign{\hrule}
width 1pt height 1pt &\omit&width .5pt&\omit&width .5pt
&\omit&width .5pt&\omit&width 1pt\cr
width1pt&cylinder&& .4424&& .4399&& .2637&width1pt\cr
\noalign{\hrule height 1pt}
}
}}
\endinsert

 The data for $r_1=100$ and $r_2=1000$ are given in Table 
3.5.
The sample size was 100,000. As the
difference between the values for the interior
arcs and those for the
exterior is in all three cases less than
$.005$, they confirm the conformal
invariance.
As a supplemental test of their reliability, we examined,
again for the model $M_0$, crossings
on a rectangle with horizontal side equal to
$122\sim A\ln(\tfrac{r_1}{r_2})$ and vertical side equal to
$332\sim 2A\pi$, $A=53$ but with periodic boundary
conditions in the vertical direction. Let $\alpha$,
$\beta$, $\gamma$ and $\delta$ be the intervals
$[y=1,y=83]$,
$[y=167,y=249]$
$[y=84,y=166]$,
and $[y=250,y=332]$
on the left side. Then we define $\pi^l_h$ as the 
probability\ of a crossing
from $\alpha$ to $\beta$ in a vertically periodic 
geometry. Two possible paths
are indicated in Figure 3.5b. We define the other
probabilities, for example $\pi^l_v$, in a similar 
fashion.

 If we extend the hypothesis
of conformal invariance to assert that
crossing probabilities on an annulus should
be equal 
(for the model $M_0$ on which we have chosen to focus)
to those
on a conformally equivalent {\it cylinder}
then, apart from
the approximations inherent in the use of finite
lattices, 
these crossing probabilities
should be equal to the corresponding probabilities
for crossings between the internal and external
intervals of the annulus. The results are also  
included in Table 3.5 as
line {\it cylinder}. The discrepancies are
larger than
$.01$
and therefore disappointing,
but tolerable especially in view of the small
inner radius $r_1$. Recall that a systematic error
$\sim 5\times 10^{-3}$ is to be expected on a square of
$200\times 200$ sites! (See \S 3.1.) In addition,
a cylinder with the given dimensions is conformally 
equivalent
to an annulus whose radii are in the ratio  $10.06$.

\fg{12 pc}\caption{Figure 3.5\rm b. Two possible paths for 
$\pi^l_{h}$.}
\endfg

 In yet another test we examined the same probabilites
for a vertical side equal to $240$ and a horizontal
side equal to $202$. This correspond to an
annulus whose outer and inner radii have a
ratio of $198.0$, and thus to
an outer radius that is virtually
infinite. The results $\hat\pi^l_h=.5003$,
$\hat\pi^l_v=.4990$, and $\hat\pi^l_{hv}=.3224$,
are,
as they should be, very close
to those that appear
in the first line of Table 3.2 and
that are familiar from experiments on the square.

 In spite of the difficulties created by the hole, it is 
nonetheless important,
especially for \S 3.7, 
to estimate, 
by simulation
and without recourse to a conformally
equivalent cylinder, crossing probabilities in exterior 
domains.
To do so we
do not use extrapolation but take advantage
in another way of conformal invariance.
For example, if we have an annulus bounded by
circles of radii $r_1$ and $r_2$ then we introduce a second
independent disk of radius $r_2$ with independent
probabilities for occupancy of the lattice points
it contains, except at the boundary. To obtain an 
admissible path 
for a given configuration
of open and closed sites we start from the inner
interval $\alpha$, or as usual from a
point in a band about $\alpha$,
but when we arrive at an open site
with a neighbor outside the larger disk, we open
the corresponding site on the second disk, and then
move as far as possible through it on open sites, allowing
ourselves to return under the same conditions to the
original annulus, and indeed to pass back and forth
between the annulus and the supplementary disk
arbitrarily many times in the effort to reach
the second interval $\beta$ on the inner boundary
of the annulus.

 Thus, in effect, we perform a roughly conformal glueing
of the annulus and the disk in order to obtain
on the Riemann sphere the exterior of the 
inner circle. With a sample of 100,000 configurations
the probability  $\pi_h$ for radii of $70$ and $350$ was
estimated to $0.5078$ and for radii of $100$ and $600$
it was found to be $0.5013$.
These values can be regarded as encouraging
confirmations of the technique, although the first is
somewhat high, differing
from $.5$ by more than our benchmark of $.005$.

\subheading{\num{3.6.} Branched percolation}
 If we apply the map $w=z^2$ to a region $D$ in the 
$z$-plane containing
the origin, then this region is realized as a branched 
covering 
of a region $D'$ in the $w$-plane. We can introduce 
crossing probabilities
for $D$ in the usual way; we can 
also lift the percolating lattice 
from $D'$ to $D$ and calculate crossing probabilities with 
respect to
it. The most general form of conformal
invariance implies that they are the same.

\fg{15.5 pc}\caption{Figure 3.6\rm a. A two-fold covering 
of the
square lattice and the image of a parallelogram by the map 
$z\to z^2$.}
\endfg

 The lattice in $D$ is best viewed, as in Figure 3.6a as a 
two-fold
covering of the lattice in $D'$, each site in $D'$, except 
those
at the branch point, being covered by two sites.
As the broken vertical line suggests, the points at the 
origin
on the two  sheets are to be identified.
The image in the $w$-plane of the the parallelogram
of aspect ratio $2.224$ and interior angles
$3\pi/8$ and $5\pi/8$ is shown in Figure 3.6b, in which 
the fine
line is the branch cut.
It is clear from the appearance of the parabolic arc,
that the upper and lower sides of the
parallelogram are horizontal. The left and right 
sides are therefore not vertical. The double covering of the
same curve appears
in Figure 3.6a. Horizontal crossings are from the sites 
marked by small squares
on one sheet to those on the second sheet. Vertical 
crossings 
are from circular
sites to circular sites. As indicated a site can
be both circular and square. 
A site is taken to be square if it is joined to a neighbor 
by a 
bond that passes through the image of
a horizontal side. The circular sites are introduced in a 
similar fashion.

\fg{24 pc}\caption{Figure 3.6\rm b. The projection of the 
image
of the parallelogram drawn in Figure 3.6a.}
\endfg

 Since the
problem is not\ in principle affected by a 
shift of the lattice, one could suppose that the
sites are never at the branch point. This is not, however,
always wise. At a branch point the effect observed in the 
previous section is even more exaggerated since the number 
$\alpha$
that appears there is now $2$, so that an indeterminacy
of $.01$ could be magnified to $.1$.  Nonetheless, to our 
astonishment,
on choosing the square lattice
such that the branch point is a site, 
which will then have eight neighbors
rather than four, we obtained simulated values
remarkably close to the true values, perhaps as a 
result of an implicit overcompensation.
The difference between simulated and predicted
values
is rarely more than $.002$.
Other choices for the branch point, for which the data are
not included, 
turned out to be far less felicitous.

The results, all for regions $D$ in the form of
parallelograms, are presented in Table 3.6, which is
self-explanatory.
Experiments
were performed for four values  $\alpha$ of the interior 
angles of the
parallelograms, and four values of the aspect ratio. The 
region
$D'$ always contained more than 200,000 sites,
occasionally many more. 
In these experiments the
second definition of the probabilities $\pi_d$
was used. (See \S 3.3.)
The probabilities $\pi_{\bar d}$
are those for a crossing between the two intervals 
complementary to those defining $\pi_d$. Thus the sum
of $\hat\pi_d$ and $\hat\pi_{\bar d}$ is expected to be 
$1$ by 
duality and universality.
Observe that the extreme
value for $\hat\pi^\square_{hv}$ is larger than
$\pi^{cft}_h$ and thus is spurious. Of course this is
not a weakness of the present experiment but a consequence
of the experiment of \S 3.2 where the values of $\hat\pi_h$
and $\hat\pi_{hv}$ for extreme values of $r$ carry important
statistical and systematic errors.

\pageinsert
\toptablecaption{\hsize=\vsize
\smc Table 3.6. \rm $\hat\pi_h,\hat\pi_v,\hat\pi_{hv},
\hat\pi_d,\hat\pi_{\bar d}$
for the image of parallelograms by the mapping $z\to z^2$.}
\vbox{
\offinterlineskip
\hrule height 1.5 pt
\halign{
\strut
\vrule width 1.5 pt\quad\hfil#\hfil\quad&
\vrule width .5pt\quad\hfil#\hfil\quad&
\vrule width .5pt\quad\hfil#\hfil\quad&
\vrule width .5pt\quad\hfil#\hfil\quad&
\vrule width .5pt\quad\hfil#\hfil\quad&
\vrule width .5pt\quad\hfil#\hfil\quad&
\vrule width .5pt\quad\hfil#\hfil\quad&
\vrule width .5pt\quad\hfil#\hfil\quad&
\vrule width .5pt\quad\hfil#\hfil\quad&
\vrule width .5pt\quad\hfil#\hfil\quad
\vrule width 1.5 pt\cr
\omit\vrule width1.5pt height.75pt&&&&&&&&&\cr
$r$&$\phi$& $\pi_h^{cft}$& $\hat\pi_h$&$\hat\pi_v$&
$\hat\pi_{hv}^{\square}$&$\hat\pi_{hv}$&
$\pi_d^{cft}$&$\hat\pi_d$&$\hat\pi_{\bar d}$\cr
\omit\vrule width 1.5pt height 1.5 pt&&&&&&&&&\cr
\noalign{\hrule height1pt }
\omit\vrule height 1 pt width 1.5pt&&&&&&&&&\cr
1.000&$%
\pi/2$&.5000&.4996&.4995&.3223&.3210&.3244&.3237&.6757\cr
1.488&$%
\pi/2$&.3002&.3018&.7018&.2600&.2615&.2939&.2972&.7065\cr
2.244&$%
\pi/2$&.1389&.1401&.8616&.1325&.1359&.2157&.2167&.7836\cr
3.309&$%
\pi/2$&.0446&.0450&.9556&.0447&.0449&.1254&.1275&.8750\cr
\omit\vrule height 1 pt width 1.5 pt&&&&&&&&&\cr
\noalign{\hrule height .5pt}
\omit\vrule height 1 pt width 1.5pt&&&&&&&&&\cr
1.000&$3%
\pi/8$&.5000&.5018&.5020&.3223&.3237&.3244&.3262&.6762\cr
1.488&$3%
\pi/8$&.2895&.2908&.7117&.2534&.2548&.2904&.2914&.7082\cr
2.224&$3%
\pi/8$&.1259&.1266&.8733&.1232&.1235&.2062&.2080&.7929\cr
3.309&$3%
\pi/8$&.0368&.0369&.9627&.0369&.0368&.1141&.1157&.8840\cr
\omit\vrule height 1 pt width1.5pt&&&&&&&&&\cr
\noalign{\hrule height .5pt}
\omit\vrule height 1 pt width 1.5pt&&&&&&&&&\cr
1.000&$%
\pi/4$&.5000&.5027&.5030&.3223&.3255&.3244&.3280&.6776\cr
1.488&$%
\pi/4$&.2491&.2500&.7522&.2262&.2269&.2754&.2760&.7260\cr
2.224&$%
\pi/4$&.0840&.0842&.9156&.0833&.0833&.1706&.1713&.8265\cr
3.309&$%
\pi/4$&.0169&.0177&.9832&.0170&.0177&.0774&.0794&.9222\cr
\omit\vrule height 1 pt width1.5 pt&&&&&&&&&\cr
\noalign{\hrule height .5pt} 
\omit\vrule height 1 pt width 1.5pt&&&&&&&&&\cr
1.000&$%
\pi/8$&.5000&.5021&.5029&.3223&.3246&.3244&.3254&.6767\cr
1.488&$%
\pi/8$&.1404&.1422&.8605&.1364&.1381&.2168&.2193&.7832\cr 
2.224&$%
\pi/8$&.0188&.0193&.9814&.0170&.0193&.0817&.0830&.9191\cr
3.309&$%
\pi/8$&.00096&.00106&.9990&.00103&.00106&.0185&.0188&.9805%
\cr
\omit\vrule height 1 pt width 1.5pt&&&&&&&&&\cr
\noalign{\hrule height 1.5pt}
}}
\vfil
\endinsert

\subheading{\num{3.7.} Percolation on compact Riemann 
surfaces}
 Any compact Riemann surface $S$ can be realized as a 
branched covering 
of the projective line $\Bbb P$, thus of $\Bbb C$ with the 
point 
at infinity added. Combining the constructions for 
unbounded domains
together with those for branched coverings, we can
introduce percolation on the surface. Various crossing 
probabilities can
be introduced. In particular each state $s$
yields a topological space, $X_s$,
formed by all bonds and the open sites and imbedded in 
$S$, and thus a homomorphism $H_1(X_s)\rightarrow H_1(S)$
from the first homology group of $X_s$ to that of $S$. 
We can ask for the probability, always at criticality, that 
a given subgroup  $Z$ of $H_1(S)$ is contained in the 
image, and expect that 
the response depends only on the conformal class of $S$. 

 We implicitly define percolation 
as being with respect to the model $M_0$
that defines the standard conformal structure on $\Bbb P$ 
and thus
on $S$. Other choices of model and conformal structure 
would be possible.
It is a matter of compatibility.

 It is possible to define a Riemann surface otherwise than 
as
a branched covering. For example, an elliptic curve can be 
obtained
as the quotient of $\Bbb C$ by a lattice $L=\Bbb Z+\Bbb 
Z\omega$, and the
percolation can be introduced directly on the surface as 
the percolation
by sites
on $aL'$ (in the limit $a\rightarrow0$) with $L'=\Bbb Z+
\Bbb Z/\sqrt{-1}$
and with (appropriately defined)
periodic boundary conditions. Conformal invariance 
implies that the probabilities on the torus
$$
     S_1=\Bbb C/(\Bbb Z+\Bbb Z\omega) \tag 3.7a
$$
and on the branched covering $S_2$ of the $x$-plane 
defined by
$$
     y^2=\prod_{i=1}^4(x-\omega_i) \tag 3.7b
$$
are the same provided the two curves are isomorphic.

 To define the percolation on (3.7a) we took $\omega=i$, 
and used
a square lattice of mesh $1\over 500$. The elliptic curve 
$S_2$ will be conformally equivalent to $S_1$ if
the points $\omega_i$ lie on the corners of a square.
We took the lattice defining the percolation to be
the usual square lattice of mesh $1$, and $\omega_i$
to be the four corners of a square with center $0$ and
sides of length $282$
parallel to the two axes. The branch cuts were
along the two horizontal lines
and were treated as for branched percolation. The 
infinite parts of the lattices on the two sheets were
handled as for exterior domains by means of a 
rough glueing along circles of radius $399$.

 The elements of $H_1(S_1)$ are naturally
labeled by pairs of integers $(m,n)$,
and $(m,n)$ and $(-m,-n)$ generate the 
same subgroup. It is easy to persuade
oneself that only primitive elements, those
for which the greatest common divisor of
$m$ and $n$ is $1$, appear as generators of subgroups
$Z$ that occur with positive probability.
We label such subgroups by a generator. Besides subgroups
with one generator, the trivial subgroup $0$ and the full
subgroup $H=H_1$ may occur.

\fighere{9pc}\caption{Figure 3.7\rm a. A possible crossing
on the elliptic curve $S_2$ for $\pi(0,1)$. (The 
``\<$\times$\<''s
are the branch points, the straight lines between them cuts;
the dashed line indicates where the conformal glueing with
another open disk takes place.)}
\fighere{9pc}\caption{Figure 3.7\rm b. A possible crossing
on the elliptic curve $S_2$ for $\pi(1,0)$.}

 We choose a conformal
equivalence of $S_1$ and $S_2$ that takes the loop around 
one of the branch cuts to the class $(0,1)$. Figure 3.7a
shows two open sets containing the branch points (crosses) 
on the upper
and lower branch of the covering. The cuts have been
chosen as indicated. The thick ellipse is one possible 
generators
of the class $(0,1)$. Figure 3.7b shows a generator for the 
class $(1,0)$. These choices
fix, in particular, an isomorphism of the two
homology groups that allows us to use
the same labels for pertinent subgroups of $H_1(S_1)$ and
$H_1(S_2)$.  The results
of our simulations are given in Table 3.7.
Rather than measure the probability $\hat \pi(0)$ that no
homology class other than $0$
occurs in the image
directly, we have for the purposes of the table  
simply {\it defined}
it to $1$ minus
the sum of the probabilities measured.
In addition to those given in the
table we measured the probabilities
for $(2,\pm1)$, $(1,\pm2)$.
Since the classes
$(1,\pm 2)$ and $(2,\pm1)$ (taken together)
appear in only $26$ of the 210000 configurations examined 
for
$S_1$ and only $16$ of the 107900 examined for $S_2$,
this definition appeared admissible. 

 We observe that the probability for
$H$ is substantially, but not intolerably, 
higher for $S_2$ than for $S_1$
and the probability for $\{0\}$ substantially lower.
The probabilities for $S_1$ are presumably closer to the
truth because they are nearly equal, as universality
and duality
demand. Moreover, simulations of $S_2$ required the use of
two devices introduced earlier: conformal glueing of
an open disk at infinity and branch points with 8 nearest 
neighbors instead of 4. As we saw, both artifices 
have limitations (See 
\S 3.5 and \S 3.6.)\ that could cause the discrepancy 
between the
value of
$\hat\pi(H)$ for $S_1$ and $S_2$,
as well as the discrepancy 
of $.024$ between  $\hat\pi(H)$ and $\hat\pi(0)$
that follows from it and our definitions.
The values of the probabilities for subgroups of rank one 
are,
however, quite close and well within the statistical errors.

%
\topinsert
\toptablecaption{\hsize=\vsize
\smc Table 3.7. \rm Probabilities of the first 
few subgroups of the\\ homology group for the two elliptic 
curves
$S_1$ and $S_2$.}
\hbox to\hsize{\hfil
\vbox{
\offinterlineskip
\hrule height 1 pt
\halign{
&\vrule#&\strut\quad\hfil#\hfil\quad\cr
width 1 pt&\omit&&$\hat\pi(H)$&&$\hat\pi(1,0)$&&$\hat%
\pi(0,1)$&&$\hat\pi(1,1)$&&
$\hat\pi(1,-1)$&&$\hat\pi(0)$&width 1 pt\cr
width 1pt height 2 pt &\omit&width .5pt&\omit&width .5pt
&\omit&width .5pt&\omit&width .5pt&\omit&width 
.5pt&\omit&width .5pt
&\omit&width 1pt\cr
\noalign{\hrule height 1 pt}
width 1pt height 2 pt &\omit&width .5pt&\omit&width .5pt
&\omit&width .5pt&\omit&width .5pt&\omit&width 
.5pt&\omit&width .5pt
&\omit&width 1pt\cr
width1pt&$S_1$&& 0.3101 && 0.1693 && 0.1686 && 0.0205 && 
0.0209 && 0.3106 &width1pt\cr
\noalign{\hrule}
width 1pt height 2 pt &\omit&width .5pt&\omit&width .5pt
&\omit&width .5pt&\omit&width .5pt&\omit&width 
.5pt&\omit&width .5pt
&\omit&width 1pt\cr
width1pt&$S_2$&& 0.3223 && 0.1700 && 0.1682 && 0.0206 && 
0.0205 && 0.2983 &width1pt\cr
\noalign{\hrule height 1 pt}
}
}
\hss}
\endinsert


\Refs
\widestnumber\key{RW2}

\ref\key{AB} \by M. Aizenman and David J. Barsky \paper 
Sharpness of the phase
transition in percolation models
\jour Comm. Math. Phys. \vol 108 \yr 1987 
\pages489--526\endref

\ref\key{BH} \by S. R.~Broadbent, J. H.~Hammersley \paper 
Percolation processes,
\RM I.~Crystals and mazes \jour Math. Proc. Cambridge 
Philos. Soc. \vol 53 \yr 1957
\pages629--641\endref

\ref\key{BPZ} \by A. A. Belavin, A. M. Polyakov, and A. B. 
Zamolodchikov
\paper Infinite conformal symmetry in
two-dimensional quantum field theory \jour
Nuclear Phys. B\vol  241 \yr1984 \pages 333--380\endref

\ref\key{C1} \by John L. Cardy \paper Conformal invariance 
and surface
critical behavior \jour Nuclear Phys. B \vol 240 \yr1984 
\pages 514--522\endref

\ref\key{C2} \bysame \paper Effect of boundary conditions 
on the operator
content of two-dimensional conformally invariant theories
\jour Nuclear Phys.  B \vol275 \yr1986 \pages200--218\endref

\ref\key{C3} \bysame \paper Boundary conditions, fusion 
rules and the Verlinde
formula \jour Nuclear Phys.  B \vol324 \yr1989 
\pages581--596\endref

\ref\key{C4} \bysame \paper Critical percolation in finite 
geometries
\jour J. Phys. A \vol25 \yr1992 \page L201\endref

\ref\key{E1} \by John W. Essam \paper Graph theory and 
statistical physics
\jour Discrete Math  \vol 1 \yr1971 \pages83--112\endref

\ref\key{E2}\bysame \paper Percolation theory \jour Rep. 
Progr. Phys. \vol 43
\yr 1980 \pages833--912\endref

\ref\key{F1} \by M. E. Fisher \paper  The theory of 
equilibrium critical
phenomena \jour Rep. Progr. Phys. \vol 30 \yr1967\pages 
615--730\endref

\ref\key{F2} \bysame 
\paper Scaling, universality and renormalization
group theory \inbook  Critical Phenomena (F. J. W.~Hahne, 
ed.), Lecture Notes
in Phys. \vol 186 \publ Springer-Verlag \publaddr New York
\yr1983 \pages 1--139\endref

\ref\key{GJ} \by James Glimm and Arthur Jaffe \book Quantum 
physics \publ Springer-Verlag \publaddr New York 
\yr1981\endref

\ref\key{G} \by G. Grimmett \book Percolation \publ 
Springer-Verlag 
\publaddr New York \yr 1989\endref

\ref\key{H} \by P. Heller \paper Experimental 
investigations of
critical phenomena \jour Rep. Progr. Phys. \vol 30 \yr1967 
\pages731--826\endref

\ref\key{K} \by H. Kesten \book Percolation theory for
mathematicians \publ Birkh\"auser \publaddr Boston
\yr 1982\endref

\ref\key{U} \by R. P. Langlands, C. Pichet, P. Pouliot, 
and Y. Saint-Aubin
\paper On the universality of crossing probabilities in 
two-dimensional
percolation \jour J. Statist. Phys. \vol 67 \yr1992 \pages 
553--574\endref

\ref\key{L} \by R. P. Langlands \paper Dualit\"at bei 
endlichen Modellen der
Perkolation \jour  Math. Nach. \vol 160 \yr1993 \pages 
7--58\endref

\ref\key{LL} \by R. P. Langlands and M.-A. Lafortune 
\paper Finite models for percolation
\paperinfo submitted for publication in the Corwin 
Memorial Volume,
Contemp. Math., Amer. Math. Soc., Providence, RI\endref

\ref\key{M} \by David S. McLachlan, Michael Blaszkiewicz, 
and Robert E. Newnham
\paper Electrical resistivity of composites
\jour J. Amer. Ceram. Soc. \vol 73 \yr1990 \pages 
2187--2203\endref

\ref\key{NF} \by D. R. Nelson and M. E. Fisher \paper 
Soluble renormalization
groups and scaling fields for the low-dimensional Ising 
systems
\jour Ann. Physics  \vol 91 \yr1975 \pages 226--274\endref

\ref\key{P} \by Jean Perrin \paper Les atomes 
\inbook Coll.~Id\'ees, vol.~222 \publ Gallimard, Paris 
\yr 1970\endref

\ref\key{RW1} \by Alvany Rocha-Caridi and Nolan Wallach 
\paper
Characters of irreducible representations of the Lie algebra
of vector fields on the circle \jour Invent. Math. \vol 72 
\yr1983 
\pages 57--75\endref

\ref\key{RW2} \bysame \paper Characters
of irreducible representations of the Virasoro algebra
\jour Math. Z.  \vol 185 \yr 1984 \pages1--21\endref

\ref\key{SA} \by Yvan Saint-Aubin \book Ph\'enom\`enes 
critiques en deux dimensions
et invariance conforme \bookinfo Course notes \publ Univ. of
Montr\'eal, 1987\endref

\ref\key{S} \by Jan V. Sengers and Anneke Levelt Sengers
\paper The critical region \inbook Chemical and 
Engineering News,
10 June 1968, pp. 104--118
\endref

\ref\key{Wo} \by Po-zen Wong \paper The statistical 
physics of 
sedimentary rock \jour Physics Today \vol 41 \yr1988 
\pages24--32\endref

\ref\key{W} \by F. Y. Wu \paper The Potts model \jour Rev. 
Modern Phys. \vol 54 
\yr1982 \pages 235--268\endref

\ref\key{Y} \by F. Yonezawa, S. Sakamoto, K. Aoki, S. 
Nos\'e, and M. Hori
\paper Percolation in Penrose tiling and its dual---in 
comparison with
analysis for Kagom\'e, dice and square lattices \jour J. 
Non-Crys.~Solids
\vol106 \yr1988 \pages 262-269\endref

\ref\key{Z} \by Robert Ziff \paper On the spanning 
probability in \RM2D percolation
\jour Phys. Rev. Lett. \vol 69 \yr1992 
\pages2670--2673\endref
\endRefs
\enddocument